\documentclass{article}
\usepackage{authblk}
\usepackage[perpage]{footmisc}
\usepackage{epsfig,endnotes,url,color,multirow,amsmath}

\newenvironment{mybullet2}{\begin{list}{$\bullet$}
    {\setlength{\topsep}{0mm}\setlength{\itemsep}{0mm}
      \setlength{\parsep}{0mm}
      \setlength{\itemindent}{0mm}\setlength{\partopsep}{0mm}
      \setlength{\labelwidth}{0mm}
      \setlength{\leftmargin}{0mm}}}{\end{list}}

\makeatletter
\def\@copyrightspace{\relax}
\makeatother
\date{}
\begin{document}

\title{A Study on the Vulnerabilities of Mobile Apps associated with Software Modules}

\author[1]{Takuya Watanabe\thanks{watanabe.takuya@lab.ntt.co.jp}}
\author[1]{Mitsuaki Akiyama}
\author[1]{Fumihiro Kanei}
\author[1]{Eitaro Shioji}
\author[1]{Yuta Takata}
\author[2]{Bo Sun}
\author[2]{Yuta Ishi}
\author[1]{Toshiki Shibahara}
\author[1]{Takeshi Yagi}
\author[2]{Tatsuya Mori\thanks{mori@nsl.cs.waseda.ac.jp}}
\affil[1]{NTT Secure Platform Laboratories}
\affil[2]{Waseda University}

\maketitle

\begin{abstract}
This paper reports a large-scale study that aims to understand how mobile application (app) vulnerabilities are associated with software libraries. We analyze {\em both free and paid apps}. Studying paid apps was quite meaningful because it helped us understand how differences in app development/maintenance affect the vulnerabilities associated with libraries. We analyzed 30k free and paid apps collected from the official Android marketplace. Our extensive analyses revealed that approximately 70\%/50\% of vulnerabilities of free/paid apps stem from software libraries, particularly from third-party libraries. Somewhat paradoxically, we found that more expensive/popular paid apps tend to have more vulnerabilities. This comes from the fact that more expensive/popular paid apps tend to have more functionality, i.e., more code and libraries, which increases the probability of vulnerabilities. Based on our findings, we provide suggestions to stakeholders of mobile app distribution ecosystems.
\end{abstract}

\section{Introduction}
\label{sec:intro}


Software libraries play a vital role in the development of modern mobile applications (app). They enable developers to improve development efficiency and app quality. In fact, Wang et al.~reported that more than 60\% of sub-packages in Android apps originate from third-party libraries~\cite{ISSTA2015_Wang}. Although software libraries offer many advantages, in some cases, they could be the source of security problems, e.g., vulnerabilities or potentially harmful functionalities. Chen et al.~\cite{SP2016_Chen} recently reported that 6.84\% of apps published to Google Play were potentially harmful apps associated with harmful software libraries.
These observations indicate that libraries can be the origins of the mobile app vulnerabilities. 

We report a large-scale study to understand how mobile app vulnerabilities are associated with software libraries. To the best of our knowledge, this is the first study that uses large datasets to systematically quantify the vulnerabilities associated with libraries. To perform our analysis, we developed two frameworks, {\em Droid-L} and {\em Droid-V}, to
detect/classify software libraries used in mobile apps and quantify how vulnerable mobile app libraries are, respectively.
By linking the output of the two frameworks, we can specify the mobile app vulnerabilities associated with libraries.
As the number of active mobile apps published in prominent mobile app marketplaces has exceeded {\em four million}~\cite{statista_app}, using a small sample of apps may result in intrinsic bias. However, analyzing all available mobile apps is not feasible. Thus, we applied proper sampling approaches to generate a dataset that is sufficient to extract statistically reliable results. We adopted two sampling approaches, i.e., top-$K$ relative to the number of installs and random sampling. Top-$K$ reflects the most influential apps and random sampling reflects the statistics of each population.

A unique and noteworthy approach of this study is that we analyze {\em both free and paid apps}. Very few studies have investigated the security of paid apps. We employ a relatively large number of paid apps to ensure statistically reliable results. Studying paid apps enables us to understand how differences in the development/maintenance of apps affect vulnerabilities associated with libraries. We examined software updates for these apps six months after they were originally collected. We collected 2M free apps to construct a database (DB) to detect/classify the libraries used in apps. In total, we used 2M free apps and 30K paid apps for our analyses.

Our primary findings are as follows.
\begin{itemize}
\item Roughly 70\% of free apps and roughly 50\% of paid apps with vulnerabilities were vulnerable due to libraries.
\item More expensive/popular paid apps tend to have more vulnerabilities than other paid apps.
\item Paid apps tend to have not been updated for longer periods than the free apps; thus, vulnerable libraries in paid apps have not been updated for longer periods than the free apps.
\item Approximately one-half of the vulnerabilities detected by existing vulnerability checking tools are found in unreachable code.
\end{itemize}

Based on these findings, we derive suggestions for stakeholders in mobile app distribution ecosystems (Section~\ref{sec:key_findings}).

The remainder of this paper is organized as follows. In Section~\ref{sec:method}, we present a high-level overview of the methodologies developed for our analysis. In Sections~\ref{sec:droid-l} and~\ref{sec:droid-v}, we describe the {\em Droid-L} and {\em Droid-V} frameworks, respectively. In Section~\ref{sec:data}, we characterize the dataset used for the analysis. Findings are presented in Section~\ref{sec:analysis}. Limitations of the analysis and future research directions are discussed in Section~\ref{sec:discuss}. We also consider ethical issues associated with this research in Section~\ref{sec:discuss}. Section~\ref{sec:related} provides a summary of related work, and conclusions are presented in Section~\ref{sec:summary}.

\begin{figure}[tbp]
  \centering
  \includegraphics[bb=0 0 1450 822,
    width=87mm,clip]{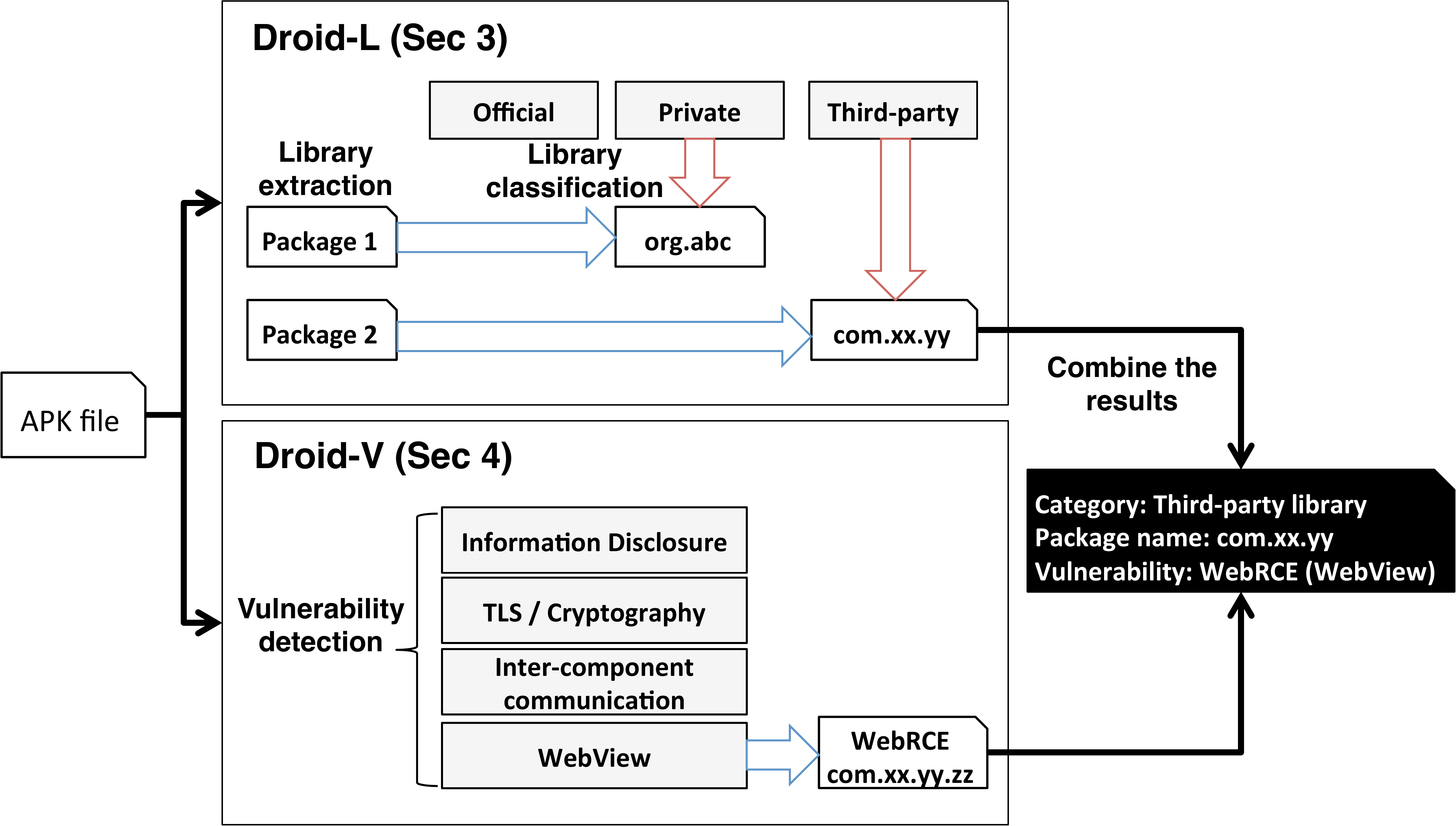}
  \caption{High-level overview of methodologies}
  \label{fig:overview}
\end{figure}

\section{Overview of Methodologies}
\label{sec:method}



Figure~\ref{fig:overview} presents a high-level overview of our approach, which consists of the {\em Droid-L} and {\em Droid-V} frameworks. {\em Droid-L} automatically detects/classifies software libraries used in mobile apps. It first extracts packages from a given APK file. In the Android OS, a {\em package} organizes multiple classes; thus, it represents the smallest unit of a software library. The package technique is used to provide modular programming in Java, which is a primary programming language for Android app development. {\em Droid-L} then classifies the extracted libraries into three primary categories. {\em Droid-V} is a compilation system that measures the degree of vulnerability of mobile app libraries. For a given APK file, we use five vulnerability checkers to detect vulnerabilities and specify Java classes and package names associated with the detected vulnerabilities. Finally, by linking {\em Droid-L} and {\em Droid-V} outputs, we can detect vulnerable libraries for a given APK file.

In the following sections, we describe {\em Droid-L} and {\em Droid-V}, and we discuss threats to the validity of each framework.

\section{Droid-L: Library detector}
\label{sec:droid-l}


The {\em Droid-L} system detects and classifies software libraries. Figure~\ref{fig:droid-l} shows an overview of the {\em Droid-L} system, which comprises a {\em fingerprint DB} and a {\em dead code checker}. For a given APK file, the system first decompiles the file and extracts packages. The system then computes a fingerprint for each package and compares the computed fingerprints to the library fingerprint DB, which we describe in the next section. The fingerprint DB returns one of three library categories, i.e., official, private, or third-party. If the DB does not return anything, this implies that the package is not a library.
Next, the system applies the dead code checker to the extracted libraries. The dead code checker employs static call graph analysis to determine if the detected library code is dead code. In the following, we describe these two components in detail.

\subsection{Building the fingerprint DB}
\label{sec:fingerprint_db}

\begin{figure}[tbp]
  \begin{center}
    \includegraphics[bb=0 0 975 439,
      width=82mm,clip]{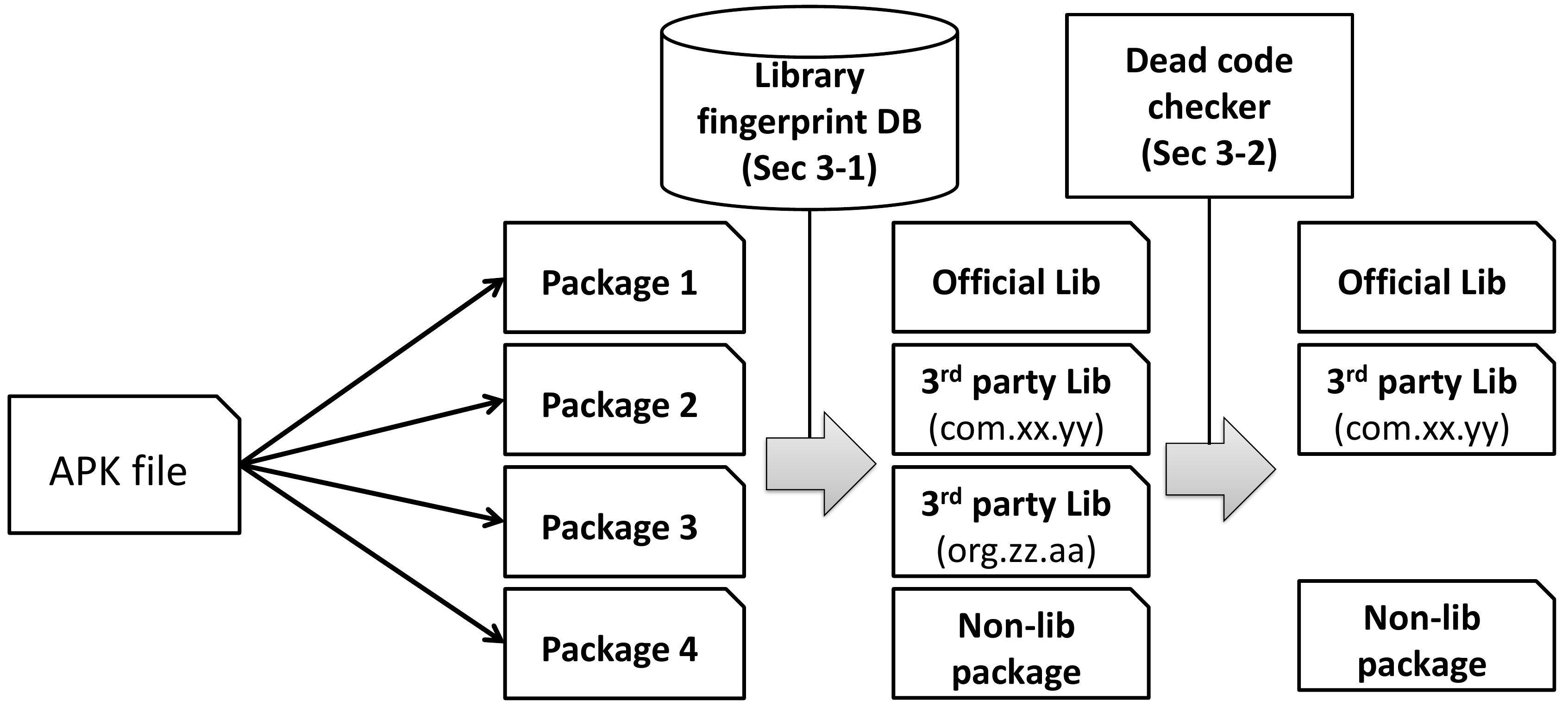}
    \caption{Overview of Droid-L.}
    \label{fig:droid-l}
  \end{center}
\end{figure}

As shown in Fig.~\ref{fig:droid-l}, the role of the fingerprint DB is to classify a given package as official, private, or third-party (Section~\ref{sec:lib_classification}). To build such a DB, we take the following two-stage approach. First, we employ cluster analysis to extract a set of packages with similar characteristics, which we call a {\em fingerprint}. A fingerprint is a unique signature that represents an extracted cluster. Then, we classify the extracted clusters using two heuristics The first heuristic is the naming convention of Java packages. Each package has an intrinsic name that may suggest which category it should belong to. For example, \url{com.google.ads} represents the AdSense library supported by the official Android SDK manager. The second heuristic is the number of distinct developer certificates per cluster. This feature is useful to determine how a detected library is used by developers. If it is a widely used public library, we will find many distinct certificates for apps that use the given library; if it is used by a single developer, the library is likely a private library.

Once we build a library fingerprint using a large collection of apps, we can extract software libraries and classify them into categories for a given app. Note that we assume that code other than the detected software libraries is attributed to app developers. We discuss the limitation of this assumption in Section~\ref{sec:droid-l-threats}.

\subsubsection{Clustering packages}
\label{sec:clustering}

To detect libraries contained in the collected apps, we begin by clustering packages. Similar packages used in many apps are clustered. A set of clustered packages possibly represents a software library. There are several ways to cluster packages~\cite{ICSE2016_Ma, ISSTA2015_Wang, SP2016_Chen}. LibRadar~\cite{ICSE2016_Ma} leverages stable API features that are resilient to code obfuscation or minor software updates. LibFinder~\cite{SP2016_Chen} compares two packages at the method level using control flow graphs.

Due to its simplicity and high scalability, we adopted the approach used in LibRadar as our base and extended it for our purpose. Note that we could adopt other clustering approaches, such as LibFinder or other clustering algorithms using features extracted from packages.


Following the LibRadar approach, we first extract packages for the given apps. Here let $p$ be an extracted package. Next, for each $p$, we derive $n(p)$, which is the total number of API calls in $p$, and $m(p)$, which is the number of distinct API calls used in $p$. Finally, for a given package $p$, we compute its fingerprint $F(p)$:
$F(p) = h(n(p), m(p))$, where $h()$ is a lightweight hash function. After processing all packages found in all apps, packages with the same fingerprint are clustered. We eliminate a cluster if it has only one package.

From the set of all package names found in a cluster, we choose the most frequently used name as the {\em representative package name (RPN)}. The RPN offers a human-interpretable representation of a cluster while removing the noise introduced by developers who modify the names of packages. While extracting RPNs is common with LibRadar, the method we use to extract
RPNs may not be identical because not all details are disclosed in Ref.~\cite{ICSE2016_Ma} and in its open-source tool~\cite{LibRadar}.
We also apply deobfuscation to package names by heuristically identifying and removing obfuscated package names (e.g., \url{zzz.a.b.c}) before choosing the RPN.
The deobfuscation is described in detail in Appendix, Section~\ref{appendix:deobfuscation}.

The extracted RPNs are useful for understanding the provenance of libraries. We use the RPNs to classify detected libraries into categories. 

\subsubsection{Library classification}
\label{sec:lib_classification}

We aim to classify detected software libraries. Note that existing library detection schemes~\cite{ICSE2016_Ma,
  ISSTA2015_Wang, SP2016_Chen} have not considered such classification. We define three library {\em categories}, i.e., official, private, and third-party, based on how they are distributed. This distinction is particularly important in relation to suggestions for managing libraries in the presence of vulnerabilities.
We use RPNs and the number of distinct certificates per library for the classification task.

The descriptions of the three categories of libraries and the ways to detect them are summarized below:

\noindent{\bf Official Libraries} are those supported by the official Android SDK Manager [2], e.g., the Android Support Library. Detected if its RPN matches one of the package names provided by the SDK Manager ,e.g., android.support.

\noindent{\bf Private Libraries} are those developed by a particular developer intended only to be used privately in apps developed by that developer, e.g., special logging/debugging libraries. Detected if all apps using the library are signed with a single signature. 

\noindent{\bf Third-Party Libraries} are those distributed freely or commercially to be used by any developers, e.g., an advertisement library. Detected if it is not classified as an official library or a private library. 

\noindent We also listed examples of RPNs for each categories in Table~\ref{tab:appen:classes} in Appendix.

Next, we classify third-party libraries into sub-categories that describe their functionality or purpose.
We considered 8 sub-categories: {\em Ad} (Advertisement), {\em Analyt} (Mobile analytics), {\em Build} (App building framework), {\em Cloud} (Cloud-based app building), {\em Dev} (Development aid), {\em Game} (Game engines), {\em Pymt} (Payment), and {\em SNS} (Social networks).
We also listed examples of RPNs for each sub-categories in Table~\ref{tab:appen:sub-category} in Appendix.
Our task is to assign a detected library/RPN to one of the categories.

First, we compile a list of package names that are associated with popular third-party libraries listed in websites such as~\cite{appbrain_lib}. Let the compiled list be ``list A.'' Second, for RPNs that are not detected in list A, we manually inspect the top package names used for at least 100 distinct apps. We summarize the results as ``list B.'' Finally, for libraries not covered by lists A and B, we apply the following prefix-matching heuristics. For a given unclassified RPN C, if there is a classified RPN D that matches a prefix of C, then C is assigned the same category as D.

Finally, using the procedures described above, we construct a fingerprint DB. Each record consists of the following three-tuple, i.e., fingerprint, deobfuscated RPN, and class/category. The fingerprint DB is employed as follows. We extract packages from a given APK file and compute a fingerprint for each package. By querying the obtained fingerprints in the DB, we can obtain corresponding deobfuscated RPNs and categories. Note that an APK file may contain code from multiple libraries in the same category, e.g., it is quite common that an app uses more than two distinct ad libraries.

\subsection{Dead code checker}
\label{sec:dead_code_checker}


Since some detected vulnerabilities may reside in dead code, we must distinguish such cases from legitimate cases. Thus, we built a dead code checker that can determine whether a given class is reachable in a generated function call tree. If all classes within a detected library are {\em not} reachable, we conclude that the detected library is dead code. The dead code checker is described in detail in Appendix, Section~\ref{appendix:Structure of dead code checker}. Note that our approach has an intrinsic limitation associated with static code analysis. This will be discussed in the next subsection.

\subsection{Threats to validity}
\label{sec:droid-l-threats}

\subsubsection{Accuracy of results}

To validate the accuracy of the results generated by the {\em Droid-L} system, we inspected the detected libraries manually. We randomly sampled 25 apps from each of four datasets, i.e., free top, free random, paid top, and paid random apps. We summarize the dataset in Section~\ref{sec:data_description}. These 100 apps contained 11,633 packages, which were grouped into 7,620 distinct clusters, and 85\% of the clusters (6,460) were detected as libraries using the fingerprint DB. The remaining packages (1,160) were not detected as libraries for the following reasons. First, the fingerprints of those libraries have been changed due to software updates. Second, some libraries use code optimization tools, such as ProGuard, which could also change fingerprints. We then inspected the 6,460 packages manually. First, we disassembled/decompiled the APK files. Then, we looked at the detected packages and inspected the classes/methods within the packages. We also searched the origins of the package source code using Internet search engines. We found that 6,308 packages (97.6\%) were classified correctly. This result clearly validates the accuracy of the {\em Droid-L} system.

\subsubsection{Dead code checker}
Static analysis, which is the basis of our approach, has the following two limitations. First, although the algorithm can exclude dead code, we cannot precisely ensure that remaining code is actually used in the app. Second, static code analysis cannot dynamically track assigned program code at run time, such as reflection. These limitations are common among static analysis approaches.

\section{Droid-V: Vulnerability checker}
\label{sec:droid-v}

Our next goal is to identify vulnerabilities in detected libraries. 
To this end, we built a vulnerability checker, i.e., {\em Droid-V}, which uses various vulnerability scanners and compiles their results for further analysis. Taking an app as input, {\em Droid-V} detects the presence of vulnerabilities and identifies where in the code the vulnerabilities reside. This information can be combined with the results of {\em Droid-L} to identify the responsible libraries. In this section, we list and describe the vulnerabilities we targeted. Some of the limitations of our system are also discussed.

\begin{table}[tbp]
  \begin{center}
    \caption{List of checked vulnerabilities}
    \label{tab:vulnlist}
    
    \begin{tabular}{l|l}
      \hline
      ID & Descriptions \\
      \hline\hline
      \multicolumn{2}{c}{Information Disclosure}\\
      \hline
      ID-GLOB & Writes data to globally accessible area \\
       ID-STOK &  Contains secret token \\
       ID-FGMT & Fragment injection vulnerability\\
      \hline
      \multicolumn{2}{c}{SSL/TLS and Cryptography}\\
      \hline
      CR-KSPW & SSL keystore is not password-protected \\
       CR-KSHC & SSL keystore is hard-coded \\
       CR-SSLV & Miscellaneous SSL validation flaws \\
       CR-CERT & Contains weak certificate \\
       CR-ECBM & ECB mode encryption is used \\
       CR-PKEY  & Contains private key \\
      \hline
      \multicolumn{2}{c}{Inter-Component Communication}\\
      \hline
      IC-CPRV & ContentProvider without export attribute \\
       IC-SRVC & Service with intent filter \\
       IC-DNGR & Declares ``dangerous'' level permission \\
       IC-EXPT & Export attribute is missing ``android:'' prefix \\
       IC-DEBG & Debuggable flag is manually set to true\\
      \hline
      \multicolumn{2}{c}{WebView}\\
      \hline
      WV-SSLV & WebView does not validate SSL \\
       WV-RCEV & WebView RCE vulnerability\\
       WV-FSYS & File system access is enabled in WebView \\
       WV-DOMS & DOM storage is enabled in WebView \\
      \hline
    \end{tabular}
  \end{center}
\end{table}

\subsection{Vulnerabilities}

As summarized in Section~\ref{sec:related}, common and influential vulnerabilities found in recent mobile platforms can be broadly classified into four categories, i.e., {\it information disclosure}, {\it SSL/TLS and cryptography}, {\it inter-component communication} (ICC), and {\it WebView}. While the first two are underlying for all softwares, not just mobile apps and devices, the last two are mobile app/device-specific issues.

Each of these vulnerability categories has the following implications. {\it Information disclosure} involves the inclusion or improper access control of sensitive information that may lead to undesired leakage. {\it Cryptography} involves the misuse of SSL/TLS and cryptographic-related code, which may lead to cryptographic integrity being compromised. {\it ICC} involves improper permissions that may allow another app to access an app's sensitive information. {\it WebView} involves the misuse of Android's WebView class, which has been a source of many vulnerabilities, including remote code execution.

Table~\ref{tab:vulnlist} lists the vulnerabilities we tested. We scanned our dataset for a total of 18
types of vulnerabilities using 2 original tools (Weak Certificate Checker and Secret Token Finder)
and 3 open source tools (AndroBugs~\cite{androbugs}, MalloDroid~\cite{mallodroid}, and QARK~\cite{qark}). 
The tools are summarized in Section~\ref{appendix:Tools} of the Appendix.

\subsection{Threats to validity}
\label{sec:droid-v-threats}
Similar to the {\em Droid-L} system, {\em Droid-V} employs static code analysis to perform a large-scale study. Clearly, static code analysis may not be able to track dynamically assigned program code. Poeplau et al.~\cite{NDSS2014_Poeplau} reported that malicious apps using dynamic code loading techniques can evade detection using offline vetting processes, e.g., static analysis, anti-virus scanning, or dynamic analysis in a sandbox without Internet connectivity. A malicious app can contain only the minimal functionality sufficient to circumvent the vetting process on Google Play, i.e., Bouncer~\cite{Bouncer}. The malicious code is downloaded only after the app is installed on a device. Dynamic code loading is an obstacle to vulnerability assessment for external code for non-malicious apps. Employing dynamic code analysis with Internet connectivity could be a promising solution to this problem. However, dynamic code analysis has several technical challenges, i.e., scalability, measuring and improving code coverage, and generating a test scenario for UI navigation~\cite{FSE2013_Machiry}. We intend to address these challenges in future work.

\section{Data}
\label{sec:data}

This section describes the free and paid app datasets used in our analysis.
Since there have been no studies that analyzed paid mobile apps on a large scale, it would be meaningful to present how they are different from free apps.
As discussed later, paid apps exhibit different characteristics compared to free apps. We construe that this reflects differences in app development and maintenance processes. We first provide an overview of datasets and then present interesting findings derived through an analysis of paid/free apps and the corresponding metadata, such as the prices of paid apps and the number of installs.

\subsection{Data description}
\label{sec:data_description}
We collected {\em paid} and {\em free} Android apps available on Google Play~\cite{googleplay}. We collected and used Android apps for two purposes. The first purpose was to generate the fingerprint DB (Section~\ref{sec:fingerprint_db}). To this end, we collected 2M free apps and 30K paid apps from Google Play. Table~\ref{tab:data_fingerprint} summarizes the data collected to generate the fingerprint DB. The second purpose was to analyze the vulnerabilities of the libraries. Collecting and analyzing all paid and free apps on Google Play was not feasible due to budgetary and labor costs; thus, we made use of filtration and sampling as follows. First, we compiled lists of paid and free apps published on Google Play. From each list, we selected both the top-$K$ and randomly sampled apps. The selected apps were divided into four sets, i.e., paid top, paid random, free top, and free random apps. The top-$K$ apps represent the most influential apps, and the randomly sampled apps reflect the statistics of each population. We used the top-5k and random-10k (rand-10k) apps for our analysis. In total, we 30k apps were used in our analyses. To further investigate the changes of vulnerabilities of apps/libraries over time, we updated these apps six months after we first collected them. Results for updated apps are presented in Section~\ref{sec:overtime}.

\begin{table}[tbp]
  \centering
  \caption{Data for building fingerprint DB}
  \label{tab:data_fingerprint}
  \begin{tabular}{l|r|r}
    \hline
    Source    & \# of APK files & Date\\
    \hline\hline
    Free  & 1,495,745      & Nov 2014\\
    Free  &   461,594      & Jun--Jul 2016\\
    Paid  &   30,000      & Jan 2016\\
    \hline
  \end{tabular}
\end{table}


\subsection{Characteristics of free/paid apps}
In the following analyses, we attempt to characterize the collected data. The derived characteristics are useful to understand the sources and impacts of vulnerabilities associated with libraries. In other words, we investigate the number of installs, prices, and number of classes.

\subsubsection{Number of installs}
\label{sec:num_installs}
\begin{figure}[tbp]
  \centering
  \includegraphics[bb=0 0 800 450,
    width=72mm,clip]{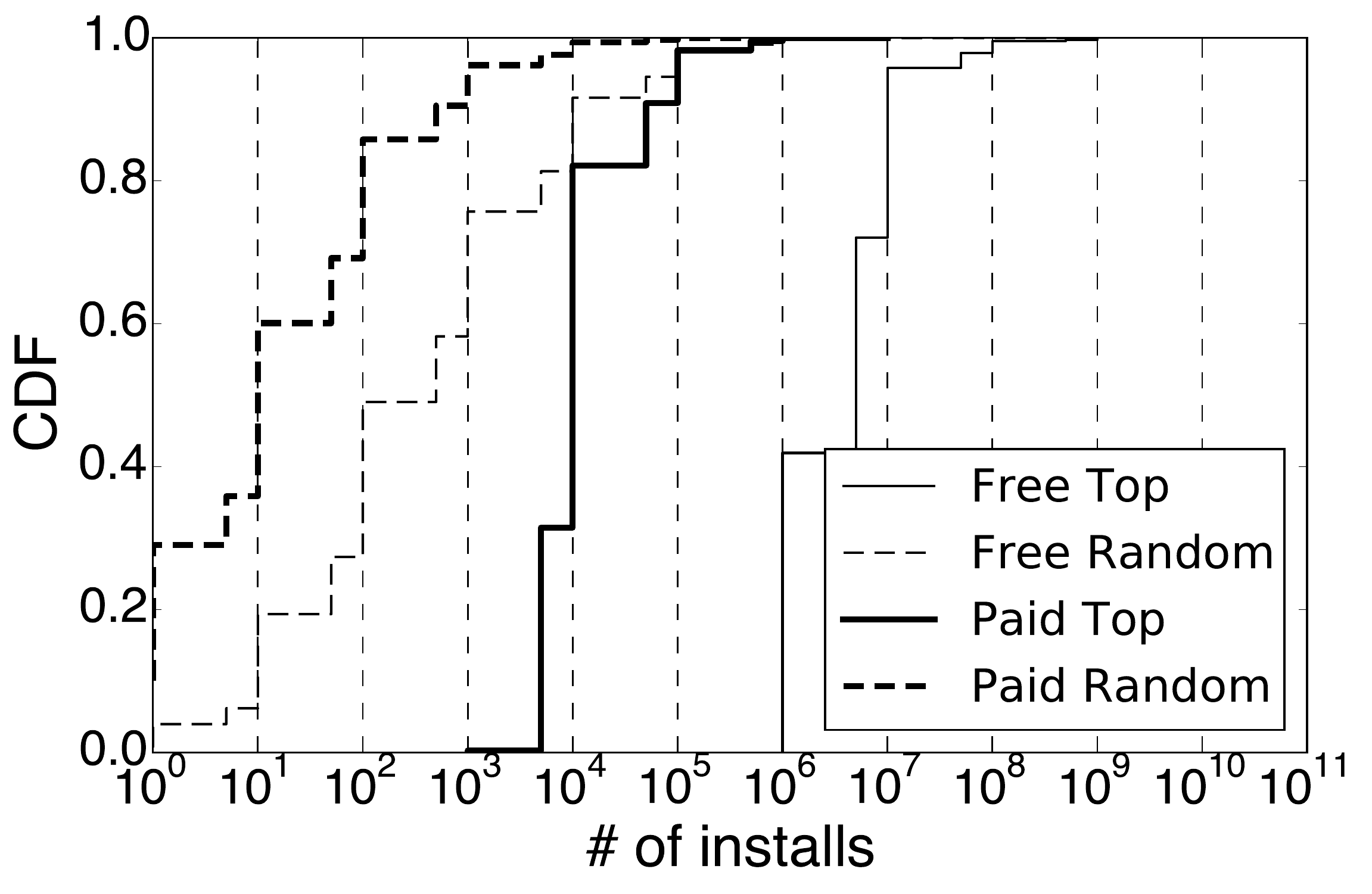}
  \caption{Distributions of number of installs.}
  \label{fig:num_installs}
\end{figure}

Figure~\ref{fig:num_installs} shows the cumulative distribution function of the number of installs per application. Note that these numbers were discretized into logarithmic ranges. Generally, free apps demonstrate a larger number of installs than paid apps. Approximately 60\% of randomly sampled paid apps show fewer than $10-50$ installs, and approximately 60\% of randomly sampled free apps show fewer than $500-1000$ installs. This tendency also applies to the top apps.

\begin{table}[tbp]
  \centering
  \caption{Statistics of prices of paid apps (USD)}
  \label{tab:price}
  \begin{tabular}{l|r|r}
    \hline
    & Top-5k   & Rand-10k \\
    \hline\hline
    mean      &      3.44&          3.30\\
    standard deviation    &       4.09&              8.90\\
    median    &     2.40&              1.51\\
    min       &      0.99&              0.99\\
    max       &      81.67&              200.0\\
    \hline
  \end{tabular}
\end{table}

\subsubsection{Prices}
\label{sec:prices}


In this study, we are interested in how  {\em prices} correlates to vulnerabilities. It is known that customers use {\em price-perceived quality heuristics}~\cite{JMR1989_Lichtenstein} when appraising the quality of a product or service. It is natural to assume that such perception might reflect expectations regarding security risks. In other words, customers may believe that a paid app has fewer security risks than a free app. After analyzing the vulnerabilities of paid mobile apps in the next section, we return to this issue in Section~\ref{sec:discuss}.

Table~\ref{tab:price} summarizes the price statistics for the top and randomly sampled paid apps. Generally, the prices of the top apps were slightly higher than those of randomly sampled apps. In addition, among random paid apps, several apps had the maximum price that can be set, i.e., 200 USD. We investigated such apps and found that most were a type of joke app, such as the ``I am rich'' app, which does not have any practical function.

\subsubsection{Time of last update}
\label{sec:update_freq}
We look into the time of the last update, which represents whether a particular app is actively developed/maintained. This information is useful for predicting the security awareness of a developer. For instance, if an app has not been updated for a long time, it may have more security risks than apps with recent updates. Ideally, it is good to use the full history of updates to measure the average time between updates. However, such information is not accessible from the web interface of the official market. As a substitute, we made use of the last date an app received an update. Although the last update date is more coarse-grained than the full history of updates, it gives us useful information about an app's development activity.

\begin{figure}[tbp]
  \centering
  \includegraphics[bb=0 0 800 420,
    width=72mm,clip]{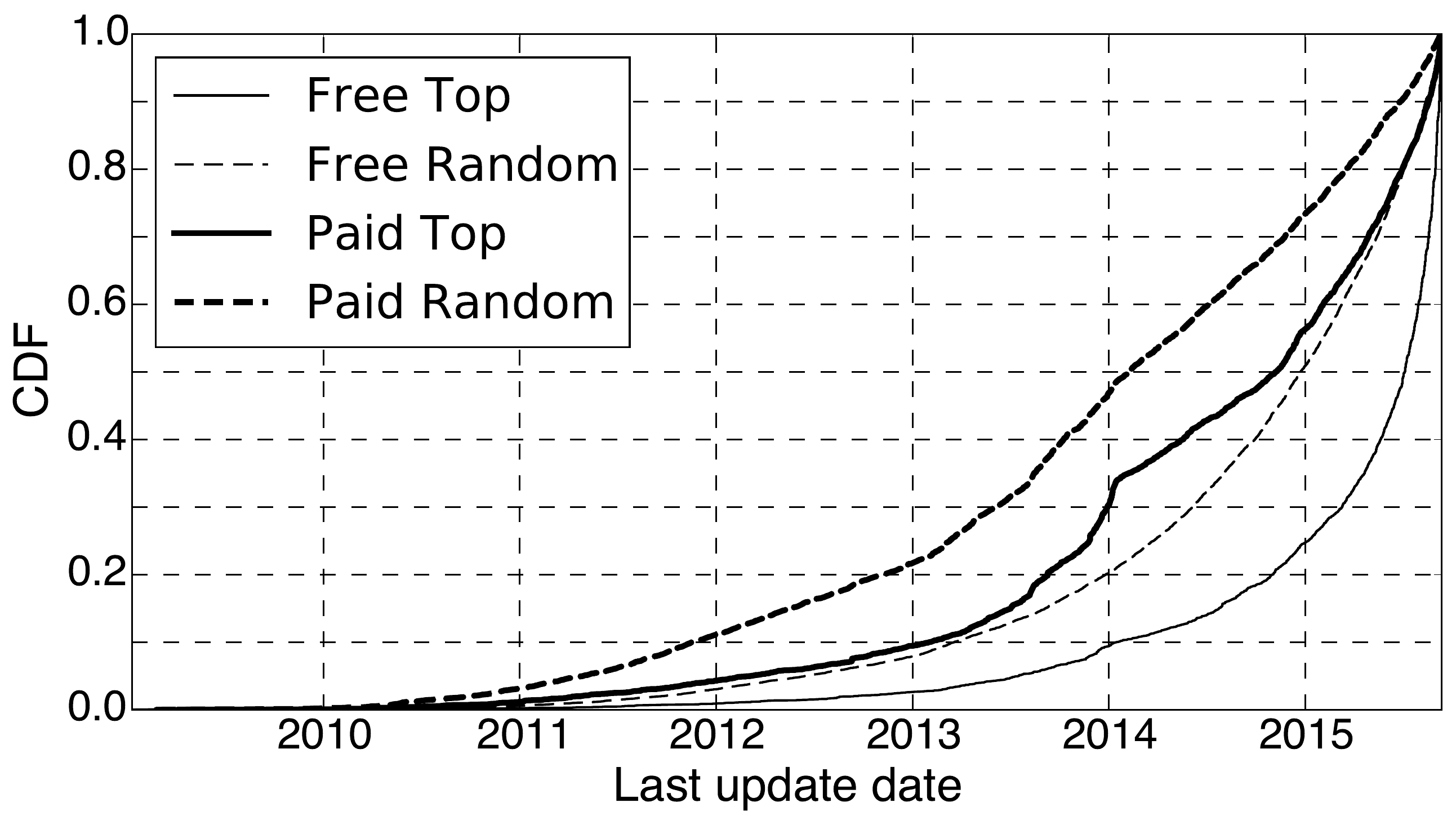}
  \caption{Distributions of last updated date}
  \label{fig:latest_update}
\end{figure}

Figure~\ref{fig:latest_update} shows the distributions of the last update date for each class of apps. Usually, we may expect that top apps tend to have a more recent last update date than randomly sampled apps for both free and paid classes. However, it is somewhat surprising that the top paid apps tend to have not been updated for longer periods than the random free apps. As presented in Section~\ref{sec:num_installs}, we consider that this originates from the fact that paid apps tend to have a lower number of installs, i.e., the higher the number of users, the more app updates. In addition, this tendency may reflect the ``sell-once-and-that's-it'' model of some paid apps.

\subsubsection{Number of classes}
\label{sec:num_classes}

\begin{figure}[tbp]
  \centering
  \includegraphics[bb=0 0 800 420,
    width=72mm,clip]{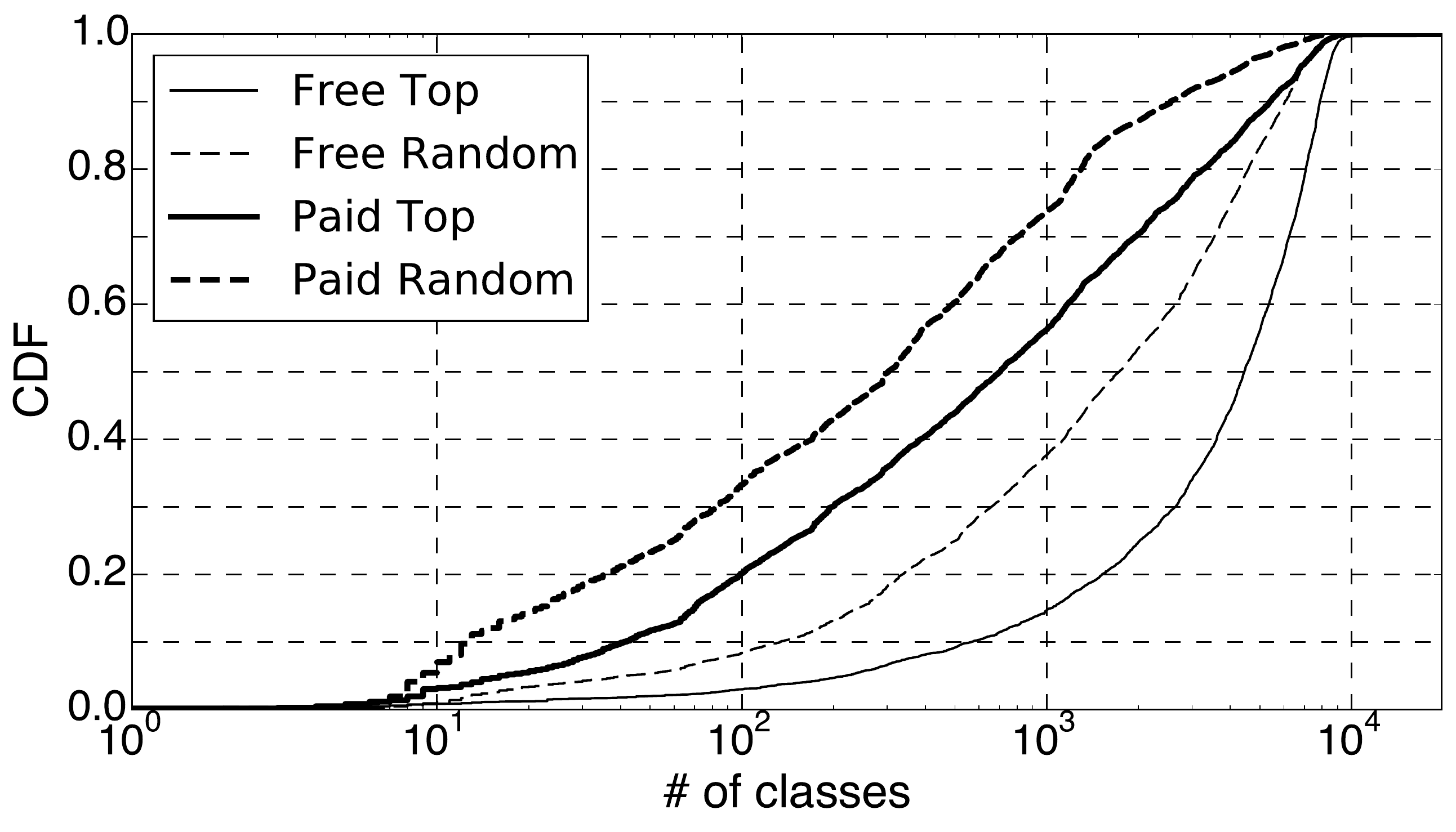}
  \caption{Distributions of number of classes}
  \label{fig:num_classes}
\end{figure}

Figure~\ref{fig:num_classes} shows the distributions of the number of classes per app. We find two general observations here. First, free apps have a larger number of classes. Second, top apps also have a larger number of classes. These observations suggest that top and free apps tend to have more functionality than random apps. As discussed in the next section, it is interesting that, for paid apps, the prices and numbers of classes exhibit a positive correlation, i.e., the more expensive an app is, the more classes (functionalities) the app has.

\section{Analysis results}
\label{sec:analysis}

This section describes the results we obtained through extensive analysis of the datasets. We first present the software libraries detected using the {\em Droid-L} system (Section~\ref{sec:libs}). Then, we present vulnerable libraries found with the two systems, i.e., {\em Droid-L} and {\em Droid-V} (Section~\ref{sec:vulns}). We also analyze how these results have changed over time (Section~\ref{sec:overtime}). Finally, we summarize the key findings derived through the analyses. Based on the findings, we provide several suggestions to stakeholders (Section~\ref{sec:key_findings}).

\subsection{Detected software libraries}
\label{sec:libs}

Table~\ref{tab:num_libs} shows statistics about the number of detected libraries per app. The results indicate two clear tendencies. First, free apps have more libraries than paid apps. Second, top apps have more libraries than randomly sampled apps. Note that this characteristic is similar to the one derived through the analysis of the number of classes (Fig.~\ref{fig:num_classes}).

\begin{table}[tbp]
  \caption{Statistics of the number of detected libraries per app.}
  \centering
  \label{tab:num_libs}
  \begin{tabular}{l|r|r|r|r}
    \hline
    Datasets & mean & std & max & min \\
    \hline\hline
    Free (Top-5k) & 10.67 & 9.39 & 101 & 0\\
    Free (Rand-10k) & 6.09 & 7.40 & 66 & 0\\
    Paid (Top-5k) & 4.86 & 6.45 & 69 & 0\\
    Paid (Rand-10k) & 3.07 & 5.20 & 74 & 0\\
    \hline
  \end{tabular}
\end{table}

\begin{table}[t]
  \centering
  \caption{Total number of detected libraries per each category.}
  \label{tab:num_libs_category}
  \begin{tabular}{l|r|r|r|r}
    \hline
    \multirow{2}{*}{Category} & Free  & Free  & Paid  &  Paid \\
                              & Top-5k & Rand-10k & Top-5k &  Rand-10k\\
    \hline\hline
    Official    & 14,404 & 21,022 & 8,977 & 11,480 \\
    Private     & 1,477  & 2,067  & 1,094  & 2,018 \\
    Third-party & 53,344 & 60,511 & 24,301 & 28,797 \\
    \hline
  \end{tabular}
  \vspace{5mm}
  \caption{Total number of detected third-party libraries per each sub-category.}
  \label{tab:num_libs_sub_category}
  \begin{tabular}{l|r|r|r|r}
    \hline
    \multirow{2}{*}{} & Free  & Free  & Paid  &  Paid \\
                              & Top-5k & Rand-10k & Top-5k &  Rand-10k\\
    \hline\hline
    Ad    &   5,453 &   2,629 &   1,122 &   884\\
    Analyt &   3,264 &   3,365 &   1,872 &   1,835\\
    Build &   269 &   1,575 &   227 &   1,084\\
    Cloud &   537 &   1,167 &   299 &   674\\
    Dev &   17,525 &   24,594 &   8,309 &   10,001\\
    Game &   2,592 &   2,419 &   1,620 &   1,736\\  
    Pymt &   831 &   1,108 &   462 &   439\\
    SNS &   2,805 &   2,560 &   1,043 &   1,020\\
    \hline
  \end{tabular}
\end{table}

Tables~\ref{tab:num_libs_category} and~\ref{tab:num_libs_sub_category} present breakdowns of the extracted libraries per category/sub-category. In Table~\ref{tab:num_libs_category}, we see that, across all the datasets, third-party libraries accounted for roughly 70\%--80\% of the detected libraries. Official libraries accounted for roughly 20\%--30\% of the detected libraries. The number of detected private libraries was much smaller than other categories. Next, in Table~\ref{tab:num_libs_sub_category}, we see that, across all datasets, development aid (Dev) was the most dominant type of third-party library. This observation was somewhat interesting to us because, before we performed the analysis, we conjectured that the most dominant type of third-party library would be advertisements. Other popular third-party libraries include advertisements (Ad), mobile analytics (Analyt), game engines (Game), and social networks (SNS).

As a result of the detection, the most popular sub-categories of the detected third-party libraries for each dataset are listed in Appendix, Table~\ref{tab:popular-3rd}.

\begin{figure}[tbp]
  \begin{center}
    \includegraphics[bb=0 0 516 298,
      width=75mm,clip]{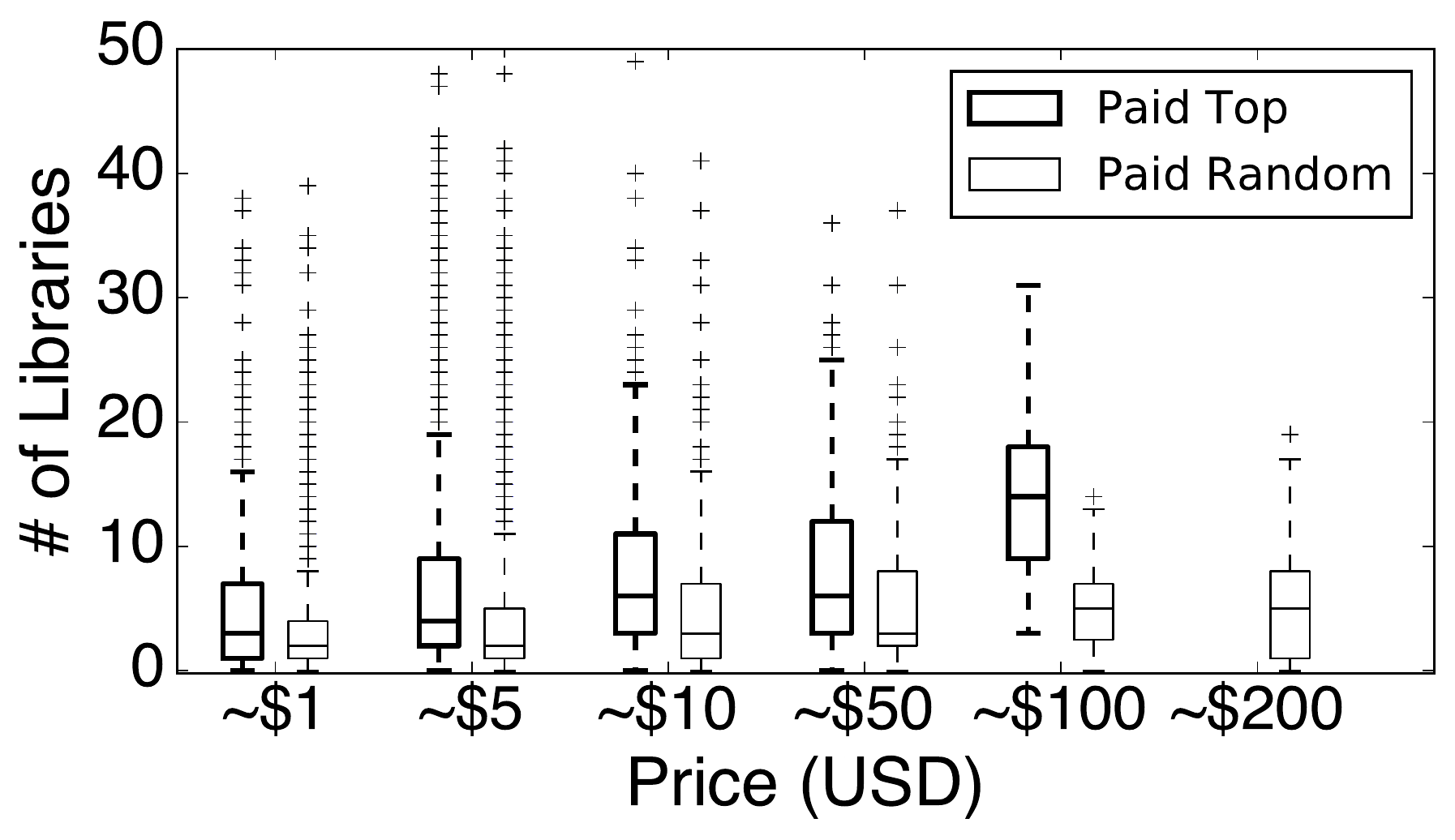}
    \caption{ Box plot of the prices of apps vs. the numbers of
      libraries in the apps.
      The top/bottom of the box is the first/third quartiles, and the
      band inside the box is the median. Whiskers represent
      the lowest/highest datum within 1.5 IQR of the first/third
      quartile where IQR is the difference between the first and third
      quartiles.
      Outliers beyond the whiskers are represented with plus
      symbols.}
    \label{fig:price_lib}
  \end{center}
\end{figure}

Finally, we inspect how the prices of apps and the number of libraries are related. Figure~\ref{fig:price_lib} presents a box plot of the price of an app against the number of libraries in the app. We make two interesting observations. First, the higher the price of an app, the more libraries the app uses. Although not conclusive, we construe that, because expensive apps tend to provide more functionality than less expensive apps, they tend to use more libraries. Second, top apps have more libraries than randomly sampled apps. Our interpretation of this finding is the same as above, i.e., top apps provide more functionality than other apps.

In the next subsection, we examine how the detected libraries are associated with vulnerabilities.

\subsection{Analysis of vulnerable apps/libraries}
\label{sec:vulns}

Here, we first present the statistics for apps that contain the summarized vulnerabilities. We then examine the libraries with vulnerabilities. We also examine how the detected vulnerable libraries changed over a period of six months. Note that an app could have a vulnerability contained in multiple libraries.

\begin{figure}[tbp]
  \begin{center}
    \includegraphics[bb=0 0 744 422,
      width=72mm,clip]{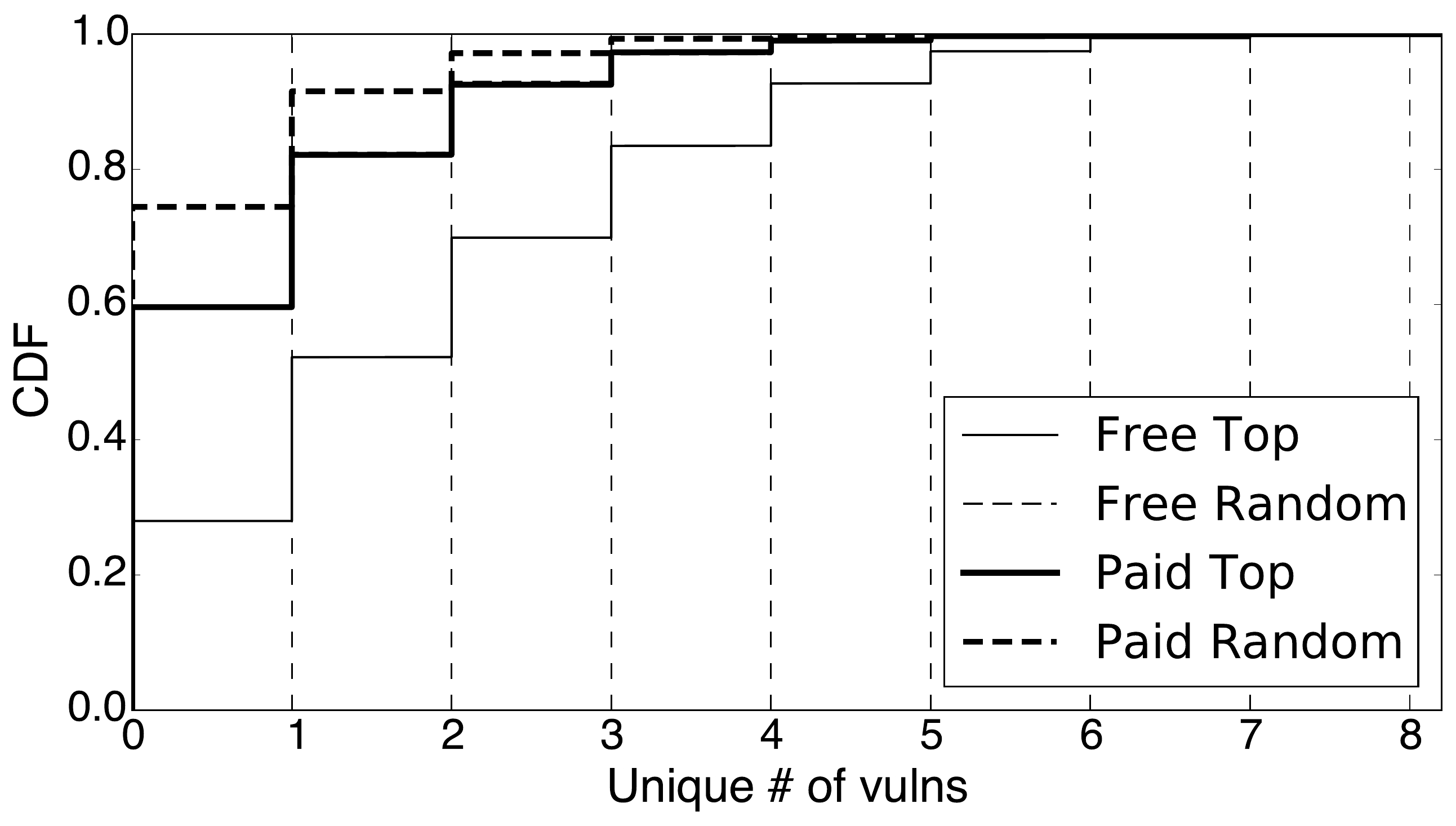}
    \caption{Distributions of the number of vulnerabilities per each
      app. Note that the lines for the Free random and Paid top overlap.}
    \label{fig:vulnerability_cdf}
  \end{center}
\end{figure}

\subsubsection{Vulnerable apps}

Figure~\ref{fig:vulnerability_cdf} shows the distributions of the number of vulnerabilities found for each app. Generally, while the number is less than that for free apps, paid apps do contain many vulnerabilities. In fact, for the top paid apps, roughly 20\% contained at least three of the vulnerabilities. We also find that top apps contain more vulnerabilities than random apps.

Table~\ref{tab:frac_dead_vuln}, $V_{dead}$ shows the fractions of detected vulnerabilities that reside in dead code. Surprisingly, approximately one-half of the vulnerabilities detected by the five independent vulnerability checkers were attributed to dead code. By combining the outputs of {\em Droid-L} and {\em Droid-V}, we can successfully exclude vulnerabilities originating from dead code. Note that we exclude dead code in the following analyses.

\begin{table}[tbp]
  \centering
  \caption{Fractions of detected vulnerabilities in dead code, $V_{dead}$,
  and fractions of apps whose vulnerabilities originated from their libraries over all the vulnerable apps, $V_{lib}$.}
  \label{tab:frac_dead_vuln}
  \begin{tabular}{l|r||r}
    \hline
    & $V_{dead}$ (\%) & $V_{lib}$ (\%) \\
    \hline\hline
    Free (Top-5k)     & 49.7 & 71.2\\
    Free (Rand-10k) & 54.7 & 71.7\\
    Paid (Top-5k)     & 40.1 & 45.9\\
    Paid (Rand-10k) & 52.3 & 52.1\\
    \hline
  \end{tabular}
\end{table}



\begin{table}[tbp]
  \centering
  \caption{Breakdown of detected vulnerabilities for each category}
  \label{tab:vuln_breakdown_category}
  \begin{tabular}{l|r|r|r|r||r}
    \hline
    &  Total   &  \multicolumn{4}{c}{fractions (\%)} \\
    \cline{3-6}
    &  \# of  & Offi- & Pri- & Third & Non\\
    &  vulns & ciai & vate & party & libs\\
    \hline\hline
    Free (Top-5k)     & 21,730 & 2.1 & 2.3 & 43.6 & 52.0 \\
    Free (Rand-10k)  & 15,516 & 1.3 & 6.4 & 59.5 & 32.8 \\
    Paid (Top-5k)     & 12,133 & 1.3 & 3.2 & 16.2 & 79.3 \\
    Paid (Rand-10k)  &  7,202 & 1.3 & 9.8 & 38.9 & 50.0\\
    \hline
  \end{tabular}
\end{table}

\subsubsection{Vulnerable libraries}

Here, we examine how many detected vulnerabilities were attributed to libraries. We also assess the origins of the vulnerable libraries. Table~\ref{tab:frac_dead_vuln}, $V_{lib}$ shows the fractions of apps whose vulnerabilities originated from their libraries for all vulnerable apps. For free apps, of the apps that contain at least one vulnerability, 71\%--72\% were vulnerable due to libraries. For paid apps, the fractions were a bit smaller; however, 46\%--52\% were vulnerable due to libraries. Thus, we conclude that most mobile apps' vulnerabilities originate from libraries.

Table~\ref{tab:frac_deadcode_vulns2}  shows the breakdown of the fractions of dead code in vulnerable and non-vulnerable libraries. The detected vulnerable libraries are more likely to contain dead code. This observation suggests that it is crucial that static vulnerability scanners include a dead code checking mechanism.

\begin{table}[tbp]
  \centering
  \caption{Fractions of dead code in the vulnerable
    libraries, $D_{v}$, and in the non-vulnerable libraries,
    $D_{n}$.}
  \label{tab:frac_deadcode_vulns2}
  \begin{tabular}{l|r|r}
    \hline
                   & $D_{v}$ (\%)  & $D_{n}$ (\%)\\
    \hline\hline
    Free (Top-5k)     & 61.1 & 34.9 \\
    Free (Rand-10k)  & 62.6 & 27.3 \\
    Paid (Top-5k)     & 71.5 & 19.5 \\
    Paid (Rand-10k)  & 66.0 & 24.5 \\
    \hline
  \end{tabular}
\end{table}

Table~\ref{tab:vuln_breakdown_category} shows a breakdown of the number of detected vulnerabilities for each category. Here, the numbers indicate the total number of Java classes that contained vulnerabilities in each set of apps. The fractions are the breakdown of the detected libraries. Note that, while this analysis counts the total number of vulnerabilities, the previous analysis shown in Table~\ref{tab:frac_dead_vuln}, $V_{lib}$ analyzed the fractions of apps with vulnerabilities due to their libraries. Free apps tend to contain more vulnerabilities in their libraries than paid apps. We also note that top apps tend to contain more vulnerabilities than random apps. These results agree with the results for app-level containment of vulnerabilities shown in Table~\ref{tab:num_libs}. Note that this also agrees with the results shown in Fig.~\ref{fig:num_classes}, i.e., more classes/libraries lead to more vulnerabilities.

Table~\ref{tab:vuln_stats} shows a breakdown of the detected vulnerabilities. While we see library-driven vulnerabilities spanning many vulnerabilities, they are particularly concentrated for the ID-GLOB, ID-FGMT, CR-KSHC, CR-SSLV, WV-SSLV, and WV-RCEV vulnerabilities. Examples of libraries that caused these vulnerabilities are IronSource (CR-KSHC), Conduit App (ID-FGMT), PayPal (CR-SSLV), Apache Cordova (WV-SSLV), and Inmobi (WV-RCEV).

\begin{table}[t]
  \caption{Breakdown of detected vulnerabilities. The numbers $X/Y$ indicate
    the total number of detected libraries ($X$) and the fractions
    (percentages) for which the vulnerabilities were due to libraries
    ($Y$). Bold fonts indicate the vulnerabilities that had a large
    impact ($>500$) and were largely contributed by libraries ($>40$ \%).}
  \label{tab:vuln_stats}
  \centering
  \small
  \begin{tabular}{l|r|r|r|r}
    \hline
    \multirow{2}{*}{Vulnerability} & Free  & Free  & Paid  &  Paid \\
                              & Top-5k & Rand-10k & Top-5k &  Rand-10k\\
    \hline\hline
    {\bf ID-GLOB} &  2166/31 &  {\bf 1469/46} &  5468/3 &  902/28 \\
    ID-STOK &  186/10 &  128/71 &  71/23 &  61/57 \\
    {\bf ID-FGMT} &  4425/18 &  {\bf 3168/49} &  2288/16 &  1362/31 \\
    CR-KSPW &  6/33 &  7/71 &  4/75 &  8/12 \\
    {\bf CR-KSHC} &  {\bf 932/60} &  485/78 &  219/44 &  124/54 \\
    {\bf CR-SSLV} &  {\bf 3644/61} &  {\bf 2733/81} &  {\bf 1195/59} &  {\bf 772/75} \\
    CR-CERT &  0/0 &  1/0 &  0/0 &  6/0 \\
    CR-ECBM &  0/0 &  0/0 &  6/16 &  9/55 \\
    CR-PKEY &  72/0 &  81/0 &  217/0 &  217/0 \\
    IC-CPRV &  237/0 &  151/0 &  164/0 &  161/0 \\
    IC-SRVC &  1167/0 &  413/0 &  533/0 &  409/0 \\
    IC-DNGR &  36/0 &  13/0 &  14/0 &  5/0 \\
    IC-EXPT &  1/0 &  1/0 &  0/0 &  0/0 \\
    IC-DEBG &  16/0 &  136/0 &  78/0 &  313/0 \\
    {\bf WV-SSLV} &  {\bf 1251/60} &  {\bf 1032/73} &  206/47 &  285/85 \\
    {\bf WV-RCEV} &  {\bf 7586/71} &  {\bf 5689/83} &  {\bf 1516/63} &  {\bf 2338/78} \\
    WV-FSYS &  3/0 &  6/0 &  141/39 &  224/54 \\
    WV-DOMS &  2/0 &  3/0 &  13/7 &  6/33 \\
    \hline
  \end{tabular}
\end{table}

\begin{figure}[tbp]
  \begin{center}
    \includegraphics[bb=0 0 822 624,
      width=87mm,clip]{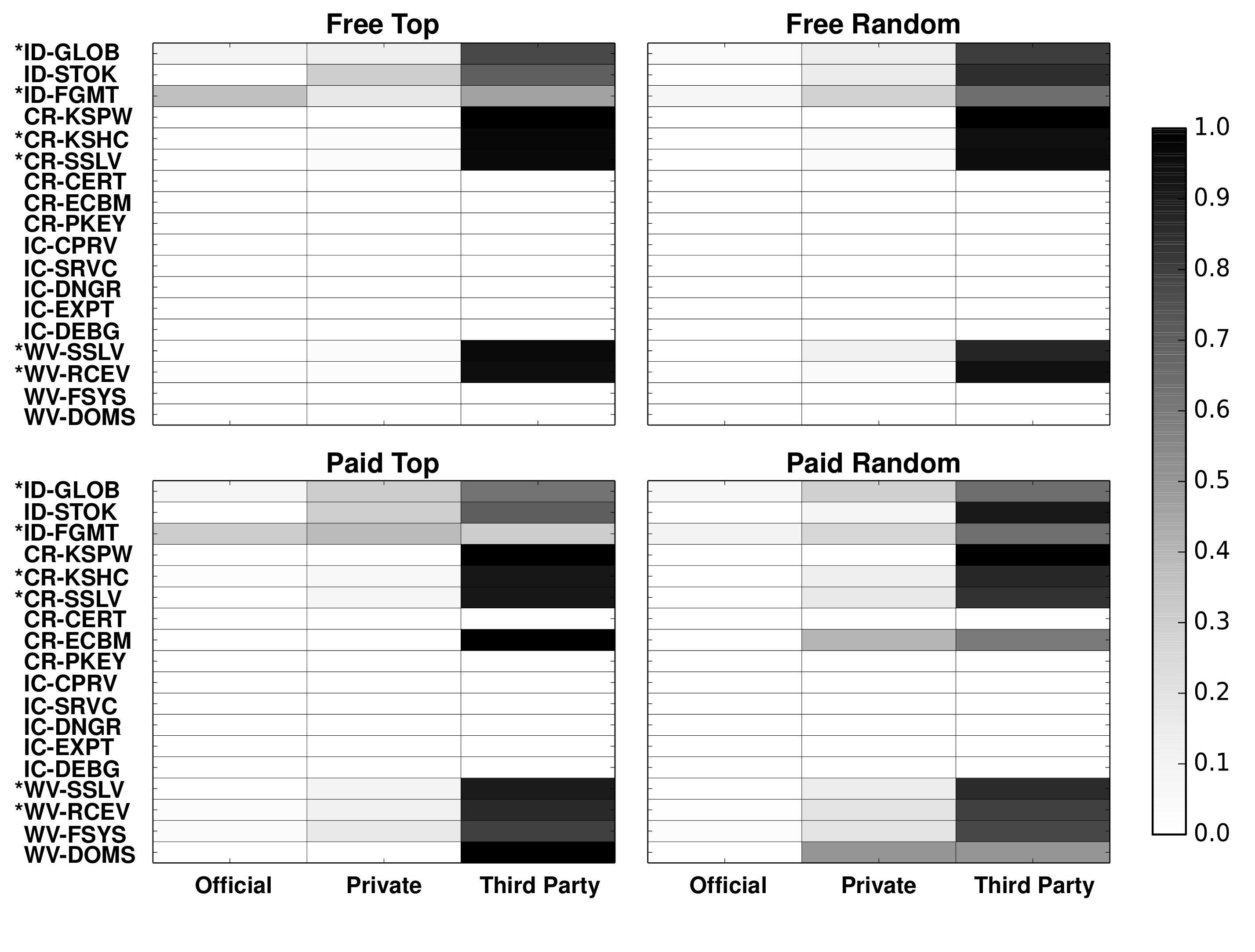}
    \includegraphics[bb=0 0 822 624,
      width=87mm,clip]{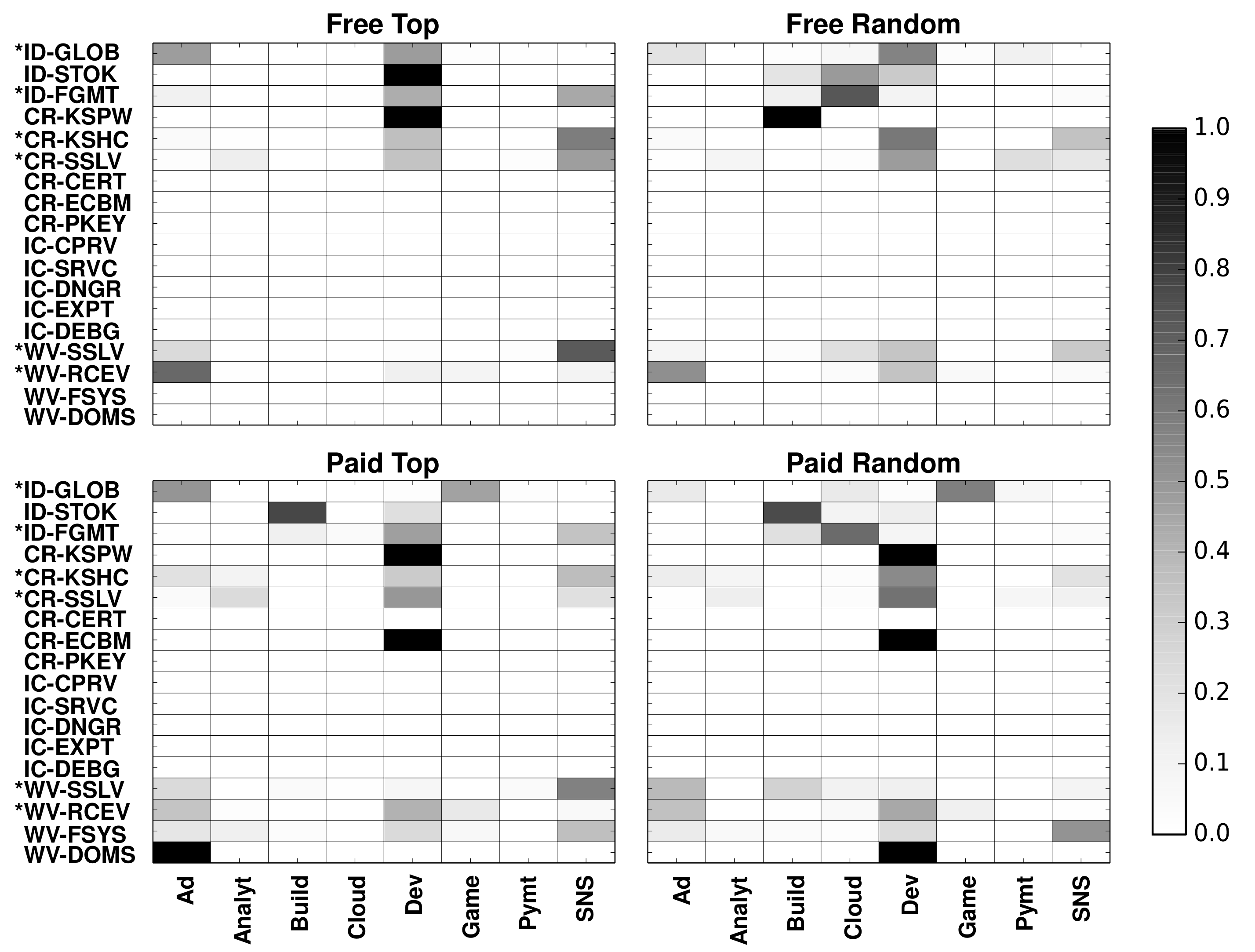}
    \caption{Relationship between library categories (top) /
      sub-categories (bottom) and vulnerabilities. Vulnerabilities shown
      with bold fonts in Table~\ref{tab:vuln_stats} are marked  with asterisk.}
      \vspace{-4mm}
    \label{fig:cat_vul}
  \end{center}
\end{figure}

The top panel of Fig.~\ref{fig:cat_vul} shows the relationship between the library categories and vulnerabilities. For each vulnerability, we inspected the distribution of categories, i.e., the fractions were normalized in each row. Most vulnerabilities were attributed to third-party libraries. In addition, although the amount was small, there are a few official libraries that contained vulnerabilities. Our manual inspection found that these vulnerabilities were attributed to certain libraries, such as Admob and the Google Mobile Service.   Thus, they are classified as ``official.'' We also found that the vulnerabilities were due to the use of older versions of libraries in which the vulnerabilities had not been fixed.

Finally, the bottom panel of Fig.~\ref{fig:cat_vul} shows the relationship between third-party library sub-categories and vulnerabilities. Among the sub-categories, Development Aid (Dev), Social Network (SNS), Advertisement (Ad), and App building framework (Build) were the main origins of the vulnerabilities. In addition, each sub-category contains intrinsic vulnerability patterns, e.g., while Ad libraries mainly contributed to the ID-GLOB and WV-RCEV, SNS libraries mainly contributed to WV-SSLV and WV-DOMS.

\subsubsection{Price vs. vulnerabilities}

\begin{figure}[tbp]
  \begin{center}
    \includegraphics[bb=0 0 520 400,
      width=75mm,clip]{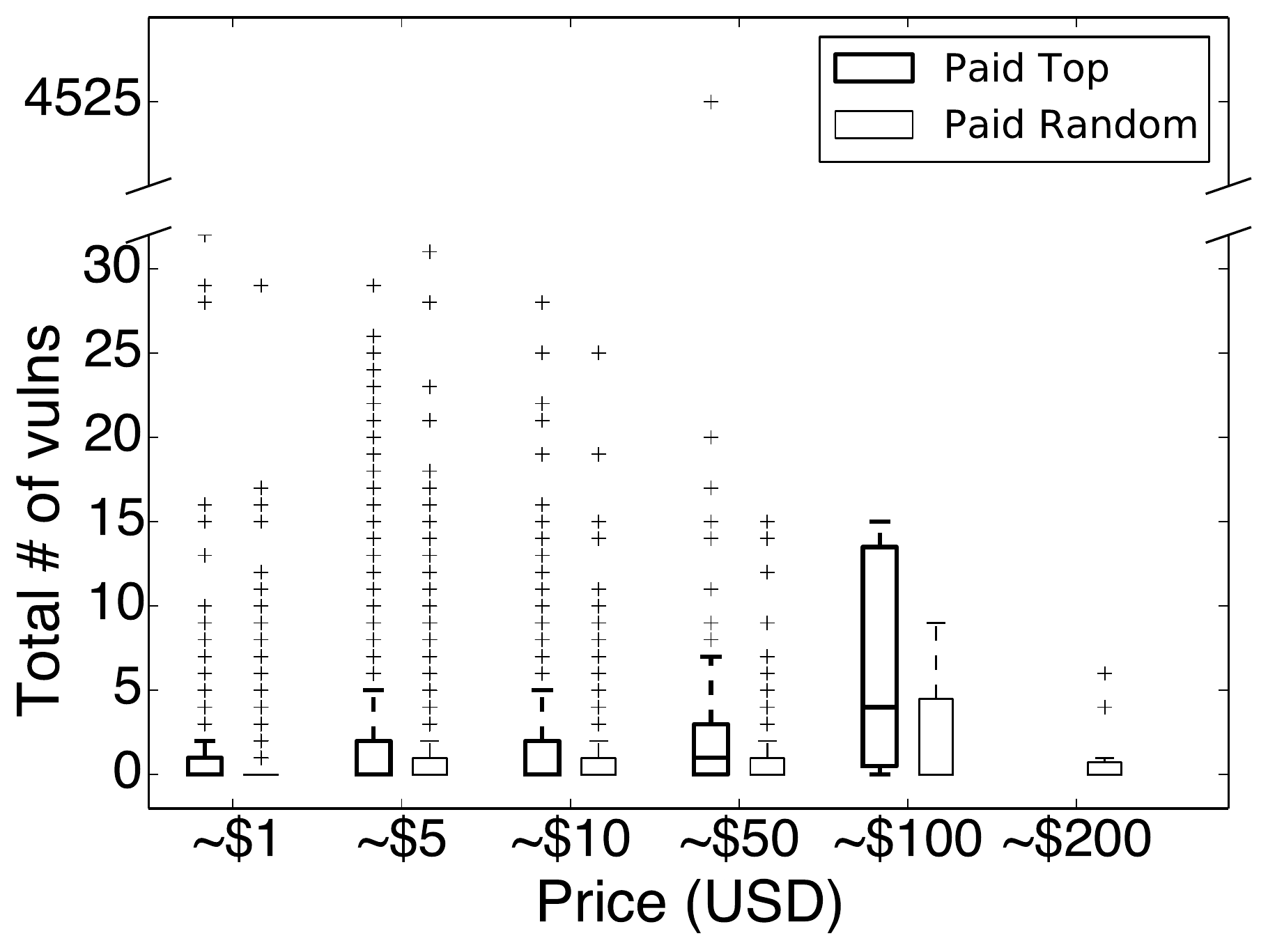}
    \caption{The prices of apps vs. the number of vulnerabilities associated with the libraries in
      the apps.}
    \label{fig:price_vulns}
  \end{center}
\end{figure}

\begin{figure}[tbp]
  \begin{center}
    \includegraphics[bb= 0 0 348 170,
      width=84mm,clip]{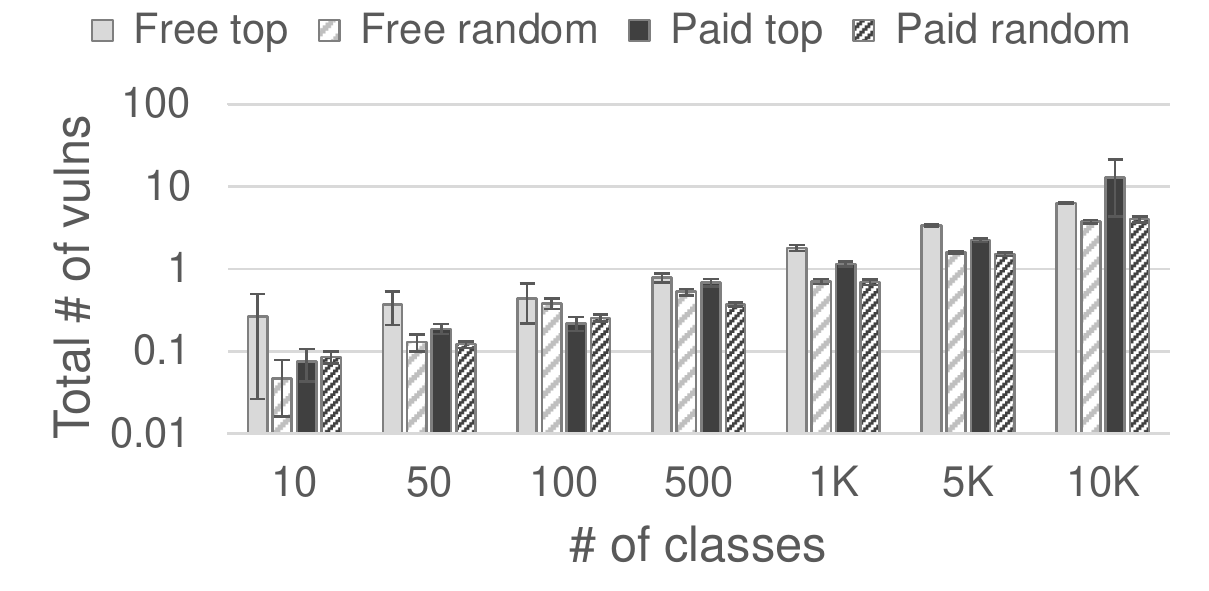}
    \caption{The number of Java classes vs. the number of vulnerabilities.}
    \label{fig:vuln_classes}
  \end{center}
\end{figure}

Figure~\ref{fig:price_vulns}  shows a correlation between the prices of paid apps and the numbers of total vulnerabilities that originated from libraries. Interestingly, more expensive apps tend to have more vulnerabilities, both in total and unique counts. In addition, top apps tend to have more vulnerabilities than random apps. This finding can be interpreted as follows. As shown in Fig.~\ref{fig:price_lib}, more expensive/popular apps tend to have more libraries. In addition, as Fig.~\ref{fig:vuln_classes} clearly shows, apps with a higher number of classes, proportional to the number of libraries, tend to have more vulnerabilities. Therefore, expensive/popular apps have more library code; thus, they are more likely to have vulnerabilities.

The anomaly shown in Fig.~\ref{fig:price_vulns}, i.e., an app in the range of 50 USD, had 4,525 total vulnerabilities. We inspect this case in detail. The app was a digital book application. The price of the app was 12 USD. All 4,525 detected vulnerabilities were attributed to a single vulnerability\footnote{{\tt MODE\_WORLD\_READABLE\_OR\_MODE\_WORLD\_WRITEABLE} which is classified as an information disclosure vulnerability}. Note that these 4,525 vulnerabilities were found in the distinct 4,525 Java classes contained in the app. For each page of the book, the app declares a unique class rather than introducing a single generic class that represents a page. In other words, every time a user turns a page of the book, the app calls a new class. To fix this vulnerability, the developer must modify all 4,525 Java classes. Despite this rather poor code implementation, it is ranked as a top paid app and had been installed more than 10,000 times at the time of data collection.

In summary, even if an app is a paid app, it is likely to have vulnerabilities. Somewhat paradoxically, more expensive/popular paid apps tend to have more vulnerabilities. These results indicate that we cannot apply {\em price-perceived quality heuristics} when we appraise the quality of an app with respect to security.

\subsection{Time-domain analysis}
\label{sec:overtime}

\begin{figure}[tbp]
  \centering
  \includegraphics[bb=0 0 504 255,
    width=86mm,clip]{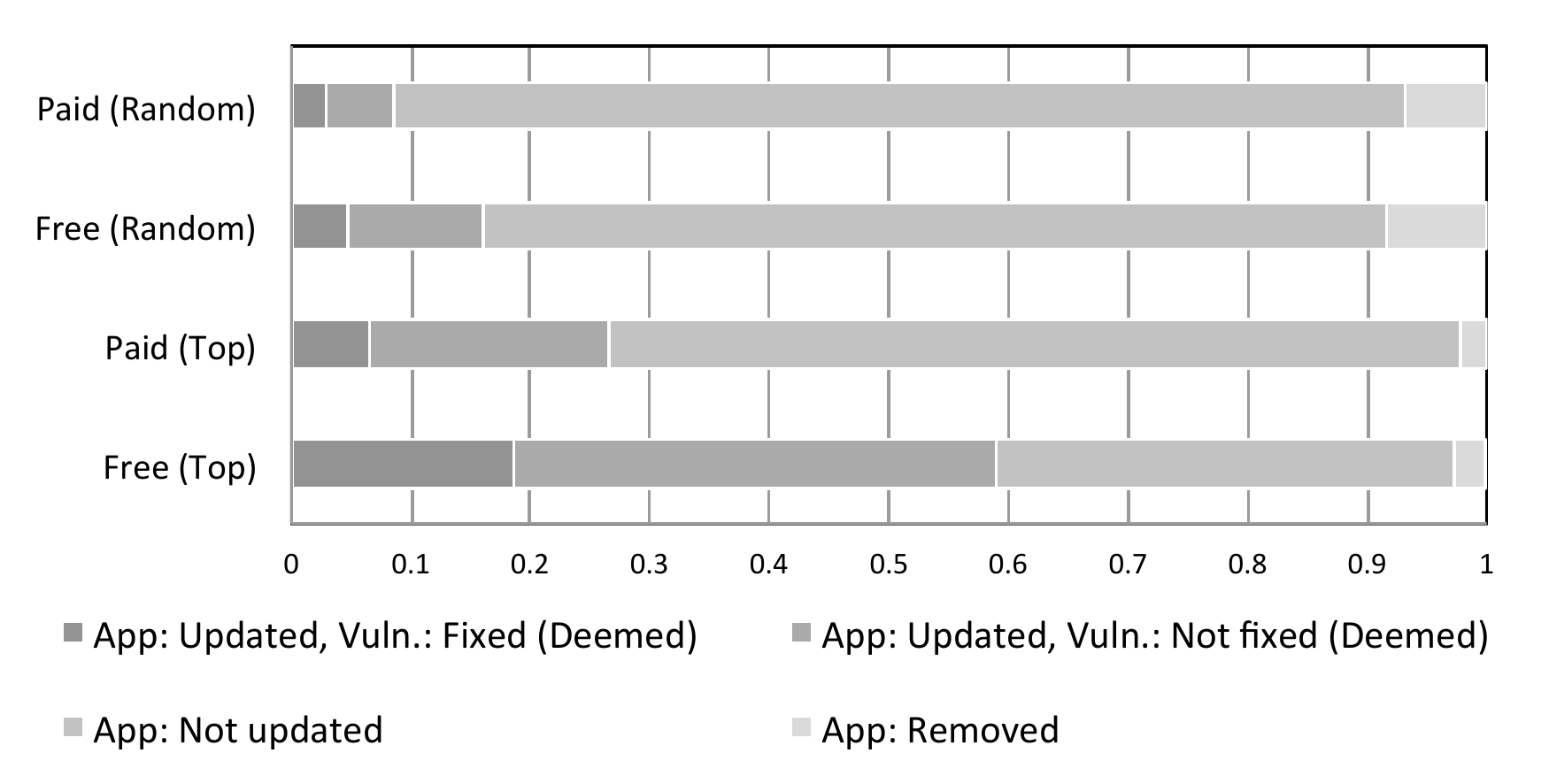}
  \caption{Statistics of apps over time}
  \label{fig:vuln_fix}
\end{figure}

We examined how vulnerabilities in apps are addressed over time. Here, we examine the status of the same apps with vulnerabilities six months after we first acquired them, and we summarize the statistics of the apps (Fig.~\ref{fig:vuln_fix}). The percentage of apps removed from the marketplace in that period was less than 9\% for each category.
The percentages of apps that were updated are as follows: free (top): 59.0\%, paid (top): 26.6\%, free (random): 16.1\%, and paid (random): 8.5\%. 
The update intervals of paid and random apps were longer than that of free and top apps. We randomly acquired over one-half of the {\it updated apps} in the four categories and confirmed which vulnerabilities were fixed using {\it Droid-V}. The percentages of apps with fixed vulnerabilities indicate the same update interval tendency, i.e., paid and random apps were more difficult to fix than free and top apps. The percentages of apps whose vulnerabilities were fixed completely are as follows: free (top): 18.5\%, paid (top): 6.4\%, free (random):4.7\%, and paid (random): 2.8\%. Unfortunately, a large proportion of apps were still vulnerable even six months after our initial investigation.

Free apps are updated in a short period due to their monetization model, i.e., updating an ad library to optimize advertising effectiveness. Therefore, vulnerabilities in libraries are fixed when apps are updated. Ruiz et al. indicated that ad libraries are frequently updated by advertising companies, and such frequent app updates force developers to update their apps~\cite{Software2014_Ruiz}. In contrast to the above {\it freemium} monetization model, {\it premium} monetization of paid apps results in less frequent updates. In addition, we assume that the effort spent on product development for random apps is less than that of top apps, and this results in infrequent updates for random apps.


The top three fixed vulnerabilities are CR-KSHC, IDSTOK, and WV-SSLV, and there is little difference between free and paid apps. The first two arise from the problem of hardcoded secret keys/tokens. WV-SSLV arises from problems with SSL validation. The reasons why these vulnerabilities are more likely fixed are as follows. First, CR-KSHC and ID-STOK are fairly easy to discover and fix. For instance, a developer can simply obfuscate secret keys/tokens. Second, since all these vulnerabilities pose a high risk to the integrity of server-side services, developers have motivation to fix them.

\subsection{Key findings and suggestions}
\label{sec:key_findings}
Here, we summarize key findings derived from our extensive analyses.
\begin{mybullet2}
\item Roughly 70\% of free apps with vulnerabilities were vulnerable due to libraries, and Roughly 50\% of paid apps with vulnerabilities were also vulnerable due to libraries.
\item Among the three library categories, third-party libraries were the main source of vulnerabilities.
\item While most vulnerable libraries originated from third-party libraries, a few official libraries were also detected as vulnerable due to the use of old versions.
\item Paid apps can contain vulnerabilities, and more expensive/popular paid apps tend to have more vulnerabilities.
\item Paid apps tend to have not been updated for longer periods than the free apps; thus, vulnerable libraries in paid apps have not been updated for longer periods than the free apps.
\item Approximately one-half of the detected vulnerabilities were attributed to dead code. We demonstrated that Droid-L can successfully exclude such cases from analysis. 
\end{mybullet2}

These key findings enable us to derive clues to remediate vulnerabilities in mobile app. We make the following suggestions to the stakeholders of mobile app distribution ecosystems, i.e., mobile app developers, mobile OS developers, app market operators, and mobile app library providers. We also offer a suggestion for the developers of vulnerability test tools.

\begin{mybullet2}
\item {\bf Mobile app developers}: Developers of apps with many classes/libraries must pay more attention to their apps. They could apply vulnerability assessment before release to at least eliminate easily-detectable vulnerabilities. 
After the release of apps, they could also check the updates of libraries they use. As we discuss in short, building a systematic update checking mechanism will be useful. 
\item {\bf Mobile OS developers}: Generally, infrequent updates lead to vulnerabilities. For instance, some paid apps adopt the ``sell-once-and-that's-it'' model. For such apps, it may not be reasonable to expect developers to perform vulnerability assessment of their products. If a mobile OS provides an automated mechanism that updates obsolete libraries/codes in an app, that could address the vulnerabilities caused by outdated software. 
\item {\bf Mobile app market operators}: Mobile app market operators should inspect all active apps using systems like {\em Droid-L} and {\em Droid-V}. In addition, they should provide vulnerability notification mechanisms that inform app developers of the sources of detected vulnerabilities. It may also be effective to present ways to update apps appropriately. Using systems like {\em Droid-L} and {\em Droid-V}, a mobile app market operator can also inform users of the potential risks of an app.
\item {\bf Mobile app library providers}: By linking {\em Droid-L} and {\em Droid-V} outputs, a list of libraries that contain vulnerabilities are generated. The results of our analysis would be useful for library providers to quickly know about the vulnerabilities and fix them. 
\item {\bf Vulnerability test developers}: As reported, roughly one-half of vulnerabilities detected by existing vulnerability check tools reside in dead code. The developers of such tools could implement a dead code checker to address this issue.
\end{mybullet2}

\section{Discussion}
\label{sec:discuss}
 
This section discusses the limitations of our analyses, user perception of security risks, and ethical issues.

\subsection{Limitations of the analyses}
As discussed in Sections~\ref{sec:droid-l-threats} and~\ref{sec:droid-v-threats}, both library detection and vulnerability checking are based on static analysis approaches. We are aware of the limitations and have described future work in previous sections. Another limitation we did not discuss is apps with {\em native code}. While our analysis focuses only on Java-written components, some Android apps contain both Java-written and native code components written in C/C++. The use of native code components is especially popular in game apps, which are required to run as quickly as possible. Afonso et al.~mentioned that ``{\it Malicious apps can use native code to hide malicious actions...}'' and surveyed how actual Android apps use native code~\cite{NDSS2016_Afonso}. They revealed that most native code components are used to improve CPU-intensive workloads, such as graphics and audio, while several hundred apps out of 1.2M contain root exploits written in native code. However, their work was not a vulnerability survey; thus, investigating vulnerabilities in native code remains a challenge.


\subsection{Ethics}
We finally discuss three ethical issues. 




{\it Acquisition of paid apps: }
All paid apps used for our analyses originated from the official Android marketplace, i.e., Google Play. We acquired all apps from the official marketplace according to the legitimate payment procedure. This means that we used our owned Google accounts to collect and purchase apps one by one without violating the Acceptable Use Policy.

{\it No additional harm: }
We conducted our app analysis in a test environment without Internet accessibility. Therefore, there was no damage to the
actual apps, devices, and services.

{\it Responsible disclosure: }
After finding new vulnerabilities in apps and libraries, we followed the principle of responsible disclosure and are now in the process of reporting them to CSIRTs and app/library developers. The disclosures will include the app and library names, the categories of vulnerability, and the source code, as well as suggested guidelines to reduce insecure code.

\section{Related Work}
\label{sec:related}

\subsection{Library analysis}

A significant amount of recent research has shed light on {\it code provenance}, which means identifying different components of an application, e.g., host apps and libraries and their developers~\cite{SP2016_Chen,ICSE2016_Ma}. These studies tackled the negative effects of a library and host app running without isolation with the same privileges. Li et al.~indicated that {\it piggybacked} apps with a library containing malicious code can mislead security analysis~\cite{Li_SANER2016}. Bhoraskar et al.~also mentioned that a host app as a whole can become vulnerable if there are bugs in the library~\cite{Security2013_Bhoraskar}.

Libraries play a vital role in improving the efficiency of developing applications and monetization (especially with ad libraries). As of 2012, 95\% of popular free Android apps contained at least one known ad library~\cite{WiSec2012_Grace}. Unfortunately, several studies have revealed the risk of an ad library automatically harvesting privacy-sensitive data without sufficient explanation to users~\cite{MoST2012_Stevens,WiSec2012_Grace}. Andow et al. analyzed popular ad libraries and identified 15 libraries as {\it madware}, which exhibits aggressive advertising behaviors~\cite{MoST2016_Andow}. Chen et al. addressed the problem of a {\it potentially harmful library (PhaLib)}, which is potentially harmful code implemented as a library, and their developed tool for finding specific code over different mobile platforms (Android and iOS) discovered 117 Android PhaLibs and 46 iOS libraries~\cite{SP2016_Chen}. To estimate the risk of information leakage, Demetriou et al. developed a tool to discover apps that expose a targeted user's privacy data to an integrated ad library~\cite{NDSS2016_Demetriou}.

Although many studies of software libraries aimed to discover malicious code, the motivation of our work is discovering vulnerabilities in libraries. To the best of our knowledge, our library analysis is the first work to classify libraries into three intrinsic categories, i.e., official, private, and third-party. This  {\it fine-grained} library analysis helps app/library developers clarify the boundaries of responsibility for countering vulnerabilities and appropriate triage countermeasures.
Backes et al.~developed a library detection method called LibScout~\cite{CCS2016_Backes}. LibScout is resilient against common code obfuscations and capable of pinpointing exact library versions. As we demonstrated in Section~\ref{sec:clustering}, the accuracy of {\em Droid-L} is high, however, we can also use these techniques as a complementary to the outputs of {\em Droid-L}.

\subsection{Vulnerability analysis}
There have been numerous studies related to vulnerability and malware/adware detection. Many of these studies applied their methods to actual apps for evaluation. Based on studies that consider vulnerabilities and threats to mobile apps and devices, we classify such vulnerabilities into four categories, i.e., {\it information disclosure}, {\it SSL/TLS and cryptography}, {\it inter-component communication}, and {\it WebView}. The first two are broad underlying issues that are not solely related to mobile apps and devices. The last two are mobile app/device-specific issues.

{\it Information disclosure}: Apps should be able to carefully process sensitive information, such as credentials; otherwise, there is a risk of information disclosure when broadcasting, logging, storing sensitive information, and setting improper file permissions. Viennot et al. conducted a survey on secret tokens for authentication embedded in app code~\cite{SIGMETRICS2014_Viennot}. Our Secret Token Finder (Section~\ref{sec:droid-v}) also finds secret tokens.

{\it SSL/TLS and cryptography}: Misuse of SSL/TLS and immature implementation of original cryptography can easily cause serious risks due to insecure communication.
Fahl et al. developed Mallodroid to find apps that misuse SSL/TLS APIs, which can result in man-in-the-middle attacks~\cite{CCS2012_Fahl}. We also used MalloDroid to find apps that misuse SSL/TLS (Section~\ref{sec:droid-v}). In addition to issues with secure communication, we have addressed weak keys used for APK certificates. Weak keys can potentially be cracked to obtain a private key, which enables the forging of a signature on a modified APK~\cite{Bockhaven_WeakKey}. Our Weak Certificate Checker discovers cryptographically weak certificates used to sign an APK file (Section~\ref{sec:droid-v}). 

{\it Inter-Component Communication} (ICC): ICC allows individual app components to be independent and enables communication between app components. Felt et al.~addressed the {\it permission re-delegation} problem, which occurs when an app with permissions performs a privileged task for an app without permissions~\cite{Security2011_Felt}. 

{\it WebView}:  WebView\footnote{iOS also provides similar classes such as UIWebView and WKWebView.} is an Android class that provides the functionalities of a custom WebKit browser to render web pages. Jin et al. revealed a new form of code injection attack using {\tt addJavascriptInterface}, which is provided by WebView and allows an app to add a bridge between JavaScript and native Java code~\cite{CCS2014_Jin}. Mutchler et al. analyzed a large number of mobile web apps embedded with WebView in terms of unsafe and leaky use of browser functionality. They found that 28\% of the apps contained at least one security vulnerability~\cite{MoST2015_Mutchler}.

Our investigation of vulnerabilities (Section~\ref{sec:droid-v}) for leveraging both original tools (Weak Certificate Checker and Secret Token Finder) and free tools (AndroBugs, MalloDroid, and QARK) broadly covered the above four categories. While AndroBugs finds various types of vulnerabilities across the four categories, other tools find different vulnerabilities that are beyond the scope of AndroBugs.

\subsection{Paid app survey at market scale}
A recent survey effort conducted by Martin et al.~\cite{martin16survey} reported that the first research into the mobile marketplaces began in 2010, and, as of the end of 2015, 155 papers have been published.

In 2012, Chakradeo et al. collected 36,710 apps from Google Play and third-party marketplaces, and they proposed lightweight triage techniques for market scale analysis in 2013~\cite{Wisec2013_Chakradeo}. In 2014, Viennot et al. presented a detailed crawler architecture to acquire apps from Google Play in a scalable manner and compiled metadata corresponding to over 1.1M apps, where they downloaded free apps and the metadata of free and paid apps~\cite{SIGMETRICS2014_Viennot}.

Although the scale of the analyzed data dramatically increases year-by-year with the exponential growth of marketplaces, the analysis of mobile apps was primarily performed in the above representative market-scale studies, except for paid apps. The total number of apps available on Google Play was approximately 2 million, and approximately 10\% of these apps were paid apps as of February 2016~\cite{AppBrain_FreeAndPaid}. Although the current market share of paid apps should be considerable and paid apps serve an important role in monetization in marketplaces, in most studies conducted at the market scale, only free apps were examined. Thus, the insights obtained from such studies were implicitly confined to free apps. Therefore, the actual security aspects of paid apps have not been considered adequately.

We investigated prior studies focusing on paid apps in terms of the number of paid apps, the origin of apps (market), analyzed object, and analytical purpose (Table~\ref{tab:paidappworks}). While most studies of paid apps covered a broad range of security topics, the analyzed properties were only extracted from market-level metadata, e.g., reviews, ratings, and the number of installs. This means that such studies did not require the actual code of the apps. There have been conventional studies that analyze the code of paid apps; however, only several hundreds of paid apps at most were analyzed. Our work achieves a double-digit increase in dataset size compared to such studies. In addition, our work was accomplished using both the code information of paid apps and market information. To the best of our knowledge, this study is the first to successfully bridge the software analysis of paid apps and market analysis at a large scale and successfully make the security of paid apps understandable at a high level.

\begin{table*}[!tbp]
\begin{center}
\caption{Summary of works on paid app analysis from 2010 to 2015}
\label{tab:paidappworks}
\footnotesize
\begin{tabular}{llrlll}
\hline
Ref                                    / Year & \# paid apps & Market & Object            & Analytical purpose\\
\hline 
\cite{CCS2011_Felt}                 / 2011 & 100          & Google & Code              & To detect overprivilege\\
\cite{DIMVA2012_Hanna}             / 2012 & 2            & Google & Code              & To detect pirated apps\\
\cite{KDD2013_Fu}                     / 2013 & 171,493      & Google & Metadata            & To understand preferences\\
\cite{MIS2013_Garg}                 / 2013 & 1,223        & Apple  & Metadata            & To infer rank-demand relationships\\
\cite{QIP2014_Eric}                 / 2014 & 486          & Apple  & Metadata            & To analyze review trends \\
\cite{Wisec2015_Seneviratne} / 2015 & 234          & Google & Code              & To analyze location privacy\\
\hline
\end{tabular}
\vspace{-3mm}
\end{center}
\end{table*}

\section{Summary}
\label{sec:summary}

To establish the assessment and remediation of mobile app vulnerabilities, understanding their origins is an imperative approach. This study has focused on mobile app libraries, which constitute most of the code in mobile apps. We have attempted to understand the provenance of mobile app libraries that cause vulnerabilities, which we have classified into four major classes, i.e., information disclosure, SSL/cryptography, ICC, and WebView. By linking the outputs of {\em Droid-L} and {\em Droid-V}, we can accurately specify the vulnerable libraries contained in apps.

A unique and noteworthy approach of this study is that we used both {\em free and paid} apps for our analysis. Since paid apps have different software development and maintenance methods, compared to free apps, they exhibit a different use of libraries or software update frequencies, and these differences affect the characteristics of vulnerabilities in the apps. Our analyses using {\em Droid-L} and {\em Droid-V} revealed that most vulnerabilities in mobile apps are caused by third-party libraries. We also found that even top paid apps do have vulnerabilities in their libraries, and many have not been updated. It was somewhat surprising that more expensive/popular paid apps tend to have more vulnerabilities. Based on the findings derived through our extensive analysis, we have proposed guidelines for mobile app developers, mobile OS developers, mobile app market operators, mobile app library providers, and vulnerability test developers.


While this work addressed the fundamental research question: 
``{\em how are the vulnerabilities of mobile apps associated with libraries?}'', we can further generalize it to:
``{\em where do the vulnerabilities of mobile apps come from?}''. 
There are many research aspects that could address this question; e.g., the app development environments, the economic models of mobile app ecosystems, the sources of information for coding, and the reuse of code. An in-depth study of such research aspects is left for future work.

\bibliographystyle{abbrv}
\bibliography{./bibliography}

\newpage
\appendix
\section{Deobfuscation}
\label{appendix:deobfuscation}
We first extract words that are separated with dots from a given package name. If at least one of the words extracted is a single letter, we identify the package name as obfuscated. For example, if the package name \url{zzz.a.b.c} is given, we extract ``zzz,'' ``a,'' ``b,'' and ``c'' as words. Since the package name included three single-letter components, we detect it as obfuscated and eliminate it from the list of RPNs. Note that this simple rule may falsely eliminate legitimate package names that include a single letter. However, we found that such cases were not common in our datasets.

\section{Example RPNs}
\label{appendix:Example RPNs}
Examples of RPNs for each category and sub-category are listed in Table~\ref{tab:appen:classes} and Table~\ref{tab:appen:sub-category}.

\begin{table}[htbp]
  \caption{Example RPNs in three categories of libraries}
  \label{tab:appen:classes}
    \begin{tabular}{l|llll}
      \hline
      Category    & \multicolumn{2}{c}{Example RPNs}\\
      \hline\hline
      Official
      &
      {\tt android.support}  &
      {\tt com.google.android}\\
      &{\tt com.google.ads}  &
      \\
      \hline
      Private
      &
      {\tt kairo} & {\tt dubbeleCom}\\
      &{\tt com.touchN} & \\
      \hline
      Third-party
      &
      {\tt com.unity3d} &
      {\tt com.flurry}\\
      &{\tt twitter4j}  &
      {\tt org.apache.cordova}\\
      \hline
    \end{tabular}
    \\
    \caption{Example RPNs in sub-categories of third-party libraries}
    \label{tab:appen:sub-category}
    \begin{tabular}{l|llll}
    \hline
    Abbreviation &  \multicolumn{2}{c}{Example RPNs} \\
    \hline\hline
    Ad & 
    \tt{com.inmobi} & 
    \tt{com.chartboost} &
    \\\hline
    Analyt & 
    \tt{com.flurry} &  
    \tt{com.crashlytics}
    \\\hline
    Build &
    \tt{com.adobe.air} & 
    \tt{org.apache.cordova} &  
    \\\hline
    Cloud & 
    \tt{com.andromo} &  
    \tt{com.biznessapps} &  
    \\\hline
    Dev &
    \tt{bolts} &  
    \tt{com.google.zxing}
    \\\hline
    Game &
    \tt{com.unity3d}  & 
    \tt{com.openfeint}
    \\\hline
    Pymt &
    \tt{com.prime31} &  
    \tt{com.paypal} &  
    \\\hline
    SNS &
    \tt{com.facebook} & 
    \tt{twitter4j} &   
    \\
    \hline
  \end{tabular}
\end{table}

\section{Dead code checker}
\label{appendix:Structure of dead code checker}
\begin{figure}[tbp]
  \begin{center}
    \includegraphics[bb=0 0 401 501, width=80mm,clip]{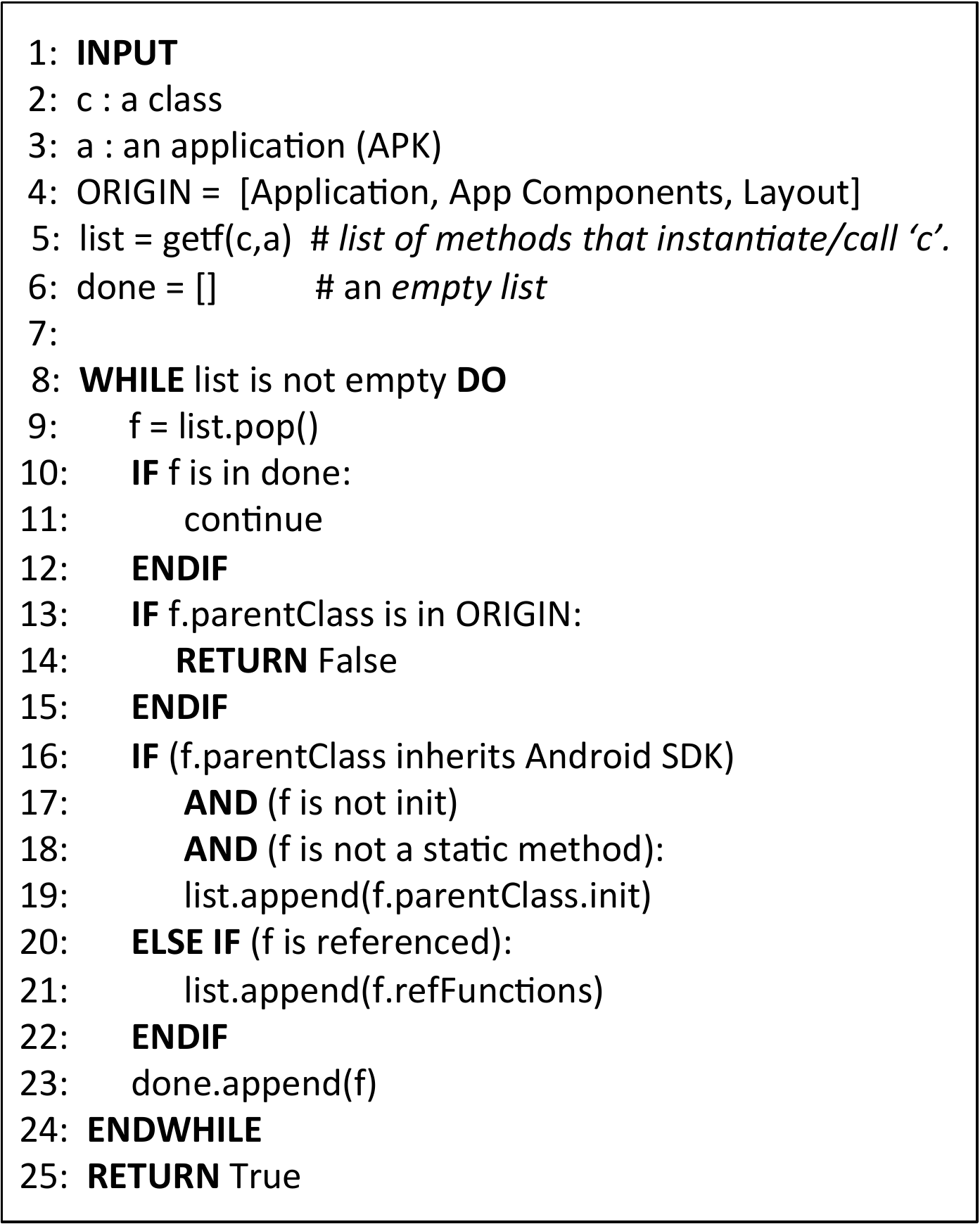}
    \caption{Pseudo code of dead code checker}
    \label{fig:cfg}
  \end{center}
\end{figure}

Figure~\ref{fig:cfg} presents the pseudo code of the dead code
checker.
For convenience, let the term {\em function} include method,
constructor execution, and field initialization; i.e., we
trace not only method calls but also class initializations.
The code checks whether a given class is a dead code (true) or not
(false). The algorithm uses depth-first search to search a function
call tree. If it finds a path from the given function to a class of
{\tt ORIGIN} (line 4), it concludes that the given class is reachable,
where {\tt ORIGIN} is composed of three classes: {\tt Application},
{\tt App Components}, and {\tt Layout}. {\tt Application }is a class
that initiates an Android app, and it is called when an app is launched. {\tt App Components} are the
essential building blocks that define the overall behavior of an
Android app, including {\tt Activities}, {\tt Services}, 
{\tt Content providers}, and {\tt Broadcast receivers}. While the 
{\tt Application} and {\tt App Components} classes need to be
specified in the manifest file of an app, the {\tt Layout} class 
does not. It is often used by ad libraries to incorporate ads using
an XML.

{\tt getf} (Line 5) is a function that returns a list of methods that
instantiate/call a class.
{\tt refFunctions} (line 21) is a function that returns a list of
functions that reference the given function. As an implementation
of {\tt refFunctions}, we adopted Androguard~\cite{androguard}, which
we modified for our purpose.
If a function of a class, say Foo, implements a function of the
Android SDK class whose code is not included in the APK, we cannot
trace the path from the function in some cases. To deal with such
cases, we made a heuristic to trace the function that calls the
init-method of class Foo (lines 16--19).
We note that the heuristics can handle several cases such as async
tasks, OS message handlers, or callbacks from framework APIs such as
{\tt onClick()}.
A method is callable if it is overridden in a subclass or an
implementation of the Android SDK and an instance of the class is
created. Async tasks, the OS message handler, or other callbacks
implement their function by overriding the methods of the Android SDK
subclass. Therefore, this should be handled by heuristics.
Finally, if there are no paths for which a given class can reach {\tt
  ORIGIN}, the algorithm concludes that the class is a dead code.

\section{Tools for vulnerability \\ checker}
\label{appendix:Tools}
In the following, we summarize the five tools we used to test the
vulnerabilities. 
Also, Table~\ref{tab:vulnlist_detail} lists the vulnerabilities we tested and the tools used for testing. 

\noindent{\bf AndroBugs}~\cite{androbugs} is an open source tool for
scanning an app for a wide variety of flaws. Its lightweight static
analysis and non-requirement of source code suits our needs
well. AndroBugs has several detection levels, but we only adopted the
``Critical'' level flaws, and in addition, we excluded those that are
not vulnerabilities, such as bugs.

\noindent{\bf Secret Token Finder} is a tool we developed to detect
secret tokens present in an app, in a way similar to how it is done in
the work of PlayDrone~\cite{SIGMETRICS2014_Viennot}. Basically, it
extracts all text strings in an APK file and searches for matches with
the regex patterns of IDs and secret tokens of known services. If such
string patterns are present in any of the strings, we marked the app
as vulnerable. We used regex patterns for AWS and Google OAuth
tokens.

\noindent{\bf MalloDroid}~\cite{mallodroid,CCS2012_Fahl} is an open-source tool for
statically analyzing an APK file for various potential SSL related
security flaws, such as the inclusion of an invalid SSL certificate or misuse
in the SSL validation logic. We considered the app vulnerable if at
least one of the flaws was detected.

\noindent{\bf Weak Certificate Checker} is a tool we implemented to
find cryptographically weak certificates used to sign an APK file. It
does several checks, such as whether a certificate was created with a
key with less than 1,024 bits, or is vulnerable to certain attacks,
e.g., Wiener's attack and common modulus attack. We considered
the app vulnerable if at least one of the flaws was detected.

\noindent{\bf QARK}~\cite{qark} is an open-source tool for analyzing 
vulnerabilities of Android apps either in source code or packaged
APKs.
The tool automates the use of multiple decompilers and combines their 
outputs to improve results. 
It covers various security related Android application vulnerabilities
such as the creation of world-readable or world-writable files,
activities that may leak data, private keys embedded in the source
code, apps that are debuggable, etc.

\begin{table*}[b]
  \begin{center}
    \caption{List of checked vulnerabilities}
    \label{tab:vulnlist_detail}
    \begin{tabular}{c|l|l|l}
      \hline
      Category & ID & CWE~\cite{CWE} & Tools \\
      \hline\hline
      \multirow{3}{*}{Information Disclosure}
      & ID-GLOB & 264 & AB \\
      \cline{2-4}
      & ID-STOK & 522 & STF \\
      \cline{2-4}
      & ID-FGMT & 264 & AB \\
      \hline
      \multirow{6}{*}{SSL/TLS and Cryptography}
      & CR-KSPW & 522 & AB,QA \\
      \cline{2-4}
      & CR-KSHC & 295, 320, 798  & AB,QA \\
      \cline{2-4}
      & CR-SSLV & 295 & MD,AB,QA \\
      \cline{2-4}
      & CR-CERT & 310, 330 & WCC,QA \\
      \cline{2-4}
      & CR-ECBM & 326, 327  & QA \\
      \cline{2-4}
      & CR-PKEY  & 312 & QA \\
      \hline
      \multirow{5}{25mm}{Inter-Component Communication}
      & IC-CPRV & 264, 926  & AB \\
      \cline{2-4}
      & IC-SRVC & 264, 285, 926  & AB \\
      \cline{2-4}
      & IC-DNGR & 264 & AB \\
      \cline{2-4}
      & IC-EXPT & 264, 926 & AB \\
      \cline{2-4}
      & IC-DEBG & 215 & QA \\
      \hline
      \multirow{4}{*}{WebView}
      & WV-SSLV & 295  & AB\\
      \cline{2-4}
      & WV-RCEV & 20, 264 & AB \\
      \cline{2-4}
      & WV-FSYS & 264  & QA \\
      \cline{2-4}
      & WV-DOMS & 264 & QA \\
      \hline
      \multicolumn{4}{c}{AB = AndroBugs, STF = Secret Token Finder,}\\
      \multicolumn{4}{c}{MD = MalloDroid, WCC = Weak Certificate Checker, and QA = QARK}
    \end{tabular}
  \end{center}
\end{table*}

\section{The most popular libraries}
Table~\ref{tab:popular-3rd} shows the most popular sub-categories of
the detected third-party libraries for each dataset.
We see that some categories have the most popular libraries in common,
e.g., {\em Facebook} is the most popular SNS third-party library across all
the datasets.
We also see some differences among the dataset.
While {\em Apache Common} was the most popular development
aid library for the paid apps, {\em Google Gson} was the most popular
development aid library for the free apps.

\begin{table}[tbp]
  \centering
  \caption{Most popular categories of detected third-party libraries.}
  \label{tab:popular-3rd}
  \begin{tabular}{l|p{14mm}|p{14mm}|p{14mm}|p{14mm}}
    \hline
    \multirow{2}{*}{}
    & Free  & Free  & Paid  &  Paid \\
      & Top-5k & Rand-10k & Top-5k &  Rand-10k\\
    \hline\hline
    \multirow{2}{*}{Ad}     & Chart Boost &  StartApp &  Chart Boost &  Inmobi\\
    \hline
    Analyt   & Flurry &  Flurry &  Flurry &  Flurry\\
    \hline
    \multirow{2}{*}{Build}  & Apache Cordova &  Apache Cordova &  Adobe Air &  Apache Cordova\\
    \hline
    \multirow{2}{*}{Cloud}  & App Inventor &  App Inventor &  App Inventor &  App Inventor\\
    \hline
    \multirow{2}{*}{Dev}    & Google Gson &  Google Gson &  Apache Common &  Apache Common\\
    \hline
    Game   & Unity3D &  Unity3D &  Unity3D &  Unity3D\\
    \hline
    Pymt   & Prime31 &  Prime31 &  Prime31 &  Prime31\\
    \hline
    SNS    & Facebook &  Facebook &  Facebook &  Facebook\\
    \hline
  \end{tabular}
\end{table}
\end{document}